\documentclass{aa}
\usepackage{graphicx}
\usepackage{txfonts}
\usepackage{longtable}
\usepackage[figuresright]{rotating}

\begin{document}

\title{The usage of Str\"omgren photometry in studies of Local Group Dwarf Spheroidal Galaxies}

\subtitle{Application to Draco: a new catalogue of Draco members and a study of the metallicity distribution function and radial gradients \thanks{Based on
observations made with the Isaac Newton Telescope, operated on the
Island of La Palma by the Isaac Newton Group in the Spanish
Observatorio del Roque de los Muchachos of the Instituto de
Astrofisica de Canarias.}$^{,}$\thanks{
Guest User, Canadian Astronomy Data Centre, which is operated by the Herzberg Institute of Astrophysics, National Research Council of Canada.}$^{,}$\thanks{Tables 2 and 6 are only available in electronic form at the CDS via anonymous ftp to cdsarc.u-strasbg.fr (130.79.128.5)}}
   \author{D. Faria \inst{1}
      \and S. Feltzing \inst{1}
      \and I. Lundstr\"om \inst{1}
      \and G. Gilmore \inst{2}
      \and G. M. Wahlgren \inst{1}
      \and A. Ardeberg \inst{1}
      \and P. Linde \inst{1}
          }

   \offprints{D. Faria}

   \institute{Lund Observatory, Lund University,
              Box 43, SE-221 00 Lund, Sweden \\ 
              \email{daniel, sofia, ingemar, glenn, arne, peter @astro.lu.se}
              \and 
              Institute of Astronomy, University of Cambridge, Madingley Road, Cambridge CB3 0HA,
United Kingdom\\
              \email{gil@ast.cam.ac.uk}
}

   \date{Received --; accepted --}

\abstract{} {In this paper we demonstrate how Str\"omgren $uvby$
     photometry can be efficiently used to: 1. Identify red giant
     branch stars that are members in a dwarf spheroidal galaxy.
     2. Derive age-independent metallicities for the same stars and
     quantify the associated errors. }
{Str\"omgren $uvby$ photometry in a 11 $\times$ 22 arcmin field
     centered on the Draco dwarf spheroidal galaxy was obtained using
     the Isaac Newton Telescope on La Palma. Members of the Draco dSph
     galaxy were identified using the surface gravity sensitive $c_1$
     index which discriminates between red giant and dwarf stars. Thus
     enabling us to distinguish the (red giant branch) members of the
     dwarf spheroidal galaxy from the foreground dwarf stars in our
     galaxy.  The method is evaluated through a comparison of our
     membership list with membership classifications in the literature
     based on radial velocities and proper motions.  The metallicity
     sensitive $m_1$ index was used to derive individual and
     age-independent metallicities for the members of the Draco dSph
     galaxy.  The derived metallicities are compared to studies based
     on high resolution spectroscopy and the agreement is found to be
     very good.}  
{We present metallicities for 169 members of the red
     giant branch in the Draco dwarf spheroidal galaxy (the largest
     sample to date). The metallicity distribution function for the
     Draco dSph galaxy shows a mean [Fe/H] = --1.74 dex with a spread
     of 0.24 dex.  The correlation between metallicity and colour for
     the stars on the red giant branch is consistent with a dominant
     old, and coeval population.  There is a possible spatial
     population gradient over the field with the most metal-rich stars
     being more centrally concentrated than the metal-poor stars.}  {}


\maketitle

\section{Introduction}
When first discovered, dwarf spheroidal galaxies (dSph) were
considered similar to the Galactic globular clusters because of their
 old stellar populations and apparent lack of gas (Shapley
1938; Hodge 1971). Over the years this picture has changed. It
is today known that dSphs show complex features like large variations
in their star formation histories and metallicities (e.g. Mateo
1998 and references therein, Shetrone et al. 2001a; Dolphin
2002). A large fraction of the dSphs also show population gradients
with a concentration of the more metal-rich stars in the central
regions (e.g. Harbeck et al. 2001).

The Draco dSph galaxy is one of the faintest companions to our galaxy,
the Milky Way, with a total luminosity of 2 $\times 10^{5} {\rm
  L_{\odot}}$ (Grillmair et al. 1998) and it lies in close proximity
to the Milky Way with a distance of $\sim$ 82 kpc (Mateo 1998).  A
large number of photometric and spectroscopic investigations have been
aimed at the Draco dSph galaxy and it is today clear that while the
star formation history shows a predominately old population (Grillmair
1998; Dolphin 2002) there exists an internal metallicity spread in the
dwarf spheroidal galaxy (e.g. Zinn 1978, 1980; Bell et al. 1985;
Carney \& Seitzer 1986; Shetrone et al. 2001a; Bellazini et al. 2002;
Winnick 2003; Cioni \& Habing 2005). Evidence for radial population
gradients similar to what is presented in Harbeck et al. (2001) for
other dSphs have also been seen in the Draco dSph galaxy in some
studies (Bellazini et al. 2002; Winnick 2003), but other studies show
no population gradients (Aparicio et al. 2001) or contradicting result
with a more centrally concentrated metal-poor population (Cioni et
al. 2005).

Recently, dSphs, and in particular the Draco dSph galaxy, have become
important tools in the study of dark matter.  Radial velocity
measurements have shown a large internal velocity dispersion leading
to M/L ratios of up to 440 (M$_\odot$/L$_\odot$)
for the Draco dSph galaxy, which would make it the most
dark-matter dominated object known (Kleyna et al. 2002; Odenkirchen et
al. 2001).  In addition, the radial velocity dispersion at large radii
shows strange behaviours. This could possibly be explained by the
presence of more than one stellar population (see discussions in
Wilkinson et al. 2004 and Mu{\~n}oz et al. 2005).  It is therefore of
great interest to further study the stellar populations in dSphs, and
in particular the Draco dSph galaxy, with respect to their metallicities and ages.

There are a number of ways to distinguish the members of a dSph from
those of our own Milky Way. The dSphs often have appreciable radial
velocities and hence measurements of the radial velocities for the
stars is a powerful, but often very time consuming, way of finding the
members. Drawbacks include binary systems (hence the stars must be
monitored for some time to resolve the binarity) and/or activity in
the atmospheres of the giant stars. Proper motions are another useful
tool. The Draco dSph galaxy has a proper motion large enough to conduct
such experiments (see Stetson 1980). A third possibility is to use a
luminosity sensitive photometric index, e.g. the $c_{\rm 1}$ index in the 
Str\"omgren photometric system, to disentangle the Red Giant Branch (RGB) stars
in the dSph from foreground dwarf stars.  This can be done with a
relatively small telescope (i.e. 2.5 m) compared to the 8 m class
telescopes needed for multi-object spectroscopy on such faint systems
as the dSphs.

 While broad-band photometric observations of dSphs are useful in
order to cover large fields and reaching faint magnitudes, the
age-metallicity degeneracy often hinders firm conclusions regarding
metallicity gradients. This is especially the case within stellar
systems with a complex star formation history and with a significant
age spread. Using spectroscopy to derive metallicities breaks the
age-metallicity degeneracy, but this is very time consuming and requires a
large telescope. The Str\"omgren $m_{\rm 1}$ index provides the
possibility to derive accurate, age-independent metallicities for RGB
stars (e.g. Hilker 2000).

In this paper we will demonstrate how Str\"omgren $uvby$ photometry
can be efficiently used to: 1. Identify members of the Draco dSph
galaxy in a field with foreground contamination from Galactic
dwarfs. 2. Obtain an age-independent metallicity distribution function
from this clean sample of members of the Draco dSph galaxy.

The article is organized as follows. In Sect.\,\ref{sect:Obs} we
describe the observations and the photometric
system. Section.\,\ref{sect:datared} deals with the data reductions
and present the colour magnitude diagram of the Draco dSph galaxy. In
Sect.\,\ref{sect:identifying} we show how the Str\"omgren $c_{\rm 1}$
index can be used to identify members of the Draco dSph galaxy and we
compare our results with results from radial velocity and proper
motion studies.  In Sect.\,\ref{sec.met.in.draco} we proceed to derive
metallicities for the members of the Draco dSph galaxy using the
Str\"omgren $m_{\rm 1}$ index. The metallicity distribution function
is presented and our results are discussed. We finally give a summary
of our results in Sect.\,\ref{sect:summary}.

\section{Observations and photometric system} 
\label{sect:Obs}

\subsection{Observations} 
The Draco dSph galaxy was observed during five nights in March 2000 at
the 2.5 m Isaac Newton Telescope (INT) on La Palma using the Wide
Field Camera (WFC). The WFC has a mosaic of four thinned AR coated EEV
4k $\times$ 2k CCDs covering a total field of 34 $\times$ 34 arcmin on
the sky with a pixel scale of 0\farcs33. The typical seeing during the
observations was $\sim$ 1\arcsec - 1\farcs4. In this paper only the
central chip will be considered.

\begin{figure}
\resizebox{\hsize}{!}{
\includegraphics{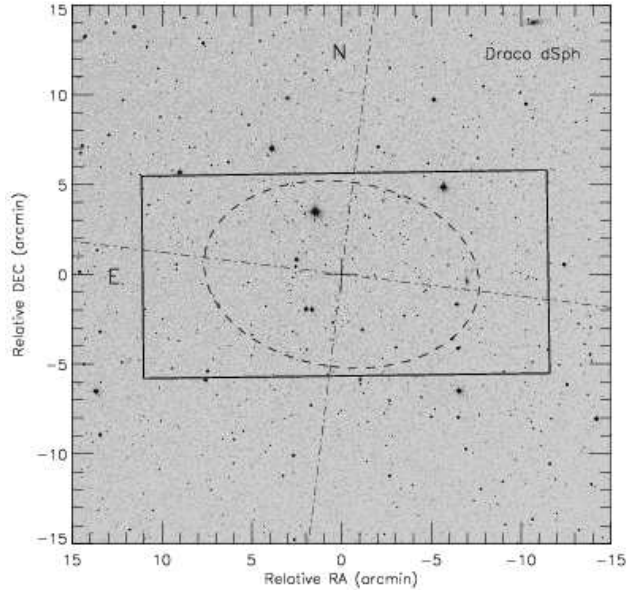}}
\caption{ESO Digital Sky Survey image centered on the Draco dSph galaxy
(RA=17h 20m 12s and DEC=+57\degr 54\arcmin 55\farcs0). The solid line
shows the observed field. The dashed ellipse shows the core radius,
r$_{core}$ = 7.7 arcmin and the dash-dotted lines the directions of the
major and minor axes. The core radius and positions for the 
major and semi-major axes are taken from Irwin \& Hatzidimitriou (1995). }
\label{fig.skymap}
\end{figure}

The observations consist of one field, centered on the Draco dSph galaxy,
RA=17h 20m 12s and DEC=+57\degr 54\arcmin 55\farcs0. Figure
\ref{fig.skymap} shows the position of the field on the sky together with
the core radius and position of the semi-major and semi-minor axes as
given by Irwin \& Hatzidimitriou (1995). The observations are
summarized in Table\,\ref{KapSou}.

\begin{table}
      \caption[]{Summary of observations. Column 1 gives the filter;
      Cols. 2-6 indicate the number of 20 minute exposures in each night,
      Col. 7 gives the total exposure time in minutes in each filter}
         \label{KapSou}
         \begin{tabular}{lllllll}
            \hline\hline
            \noalign{\smallskip}
            Filter & \multicolumn{5}{l}{Night} & Total  \\
                   & 22 & 23 & 24 & 25 & 28    &  Exp. time   \\
                   & [min]&[min]&[min]&[min]&[min]&[min]\\
            \noalign{\smallskip}
            \hline
            \noalign{\smallskip}
              $y$ & 3$\times$20 & - & 1$\times$20 & 2$\times$20 & - & 120  \\
	      $b$ & 2$\times$20 & 2$\times$20 & 2$\times$20 & - & - & 120   \\
	      $v$ & - & 5$\times$20 & - & 1$\times$20 & - & 120   \\
              $u$ & - & - & 3$\times$20 & 1$\times$20 & 4$\times$20 & 160   \\ 
            \noalign{\smallskip}
            \hline
         \end{tabular}
\end{table}

Photometric standard stars were chosen from the list in Schuster \&
Nissen (1988) of secondary Str\"omgren standard stars. The reason for
using secondary standards rather than primary Str\"omgren standard
stars is that the latter are too bright to observe with a 2.5 m
telescope. During each night approximately 15 standard stars with a
large span in magnitude and colour indices, $8.137 \leq y \leq 12.828,
0.237\leq (b-y)\leq0.611, 0.032\leq m_{1}\leq0.610,$ and $0.094\leq
c_{1}\leq0.490$, were observed at an airmass around 1.3. In total 43
standard stars were observed. These observations were used to obtain
the colour terms in the calibration.  To derive the atmospheric
extinction for each night we also observed two standard stars at
airmasses ranging from 1 to $\sim$2.2. These two stars will hereafter
be referred to as the two extinction stars. Normally each of them was
observed at six different airmasses. Table \,\ref{tab.standards} lists
the photometric data used in our calibration of the observed counts
onto the standard system.

\subsection{Photometric system}
\label{sect:photsystem}
Since the standard Str\"omgren system for giants is based on data that
does not contain any significant number of metal-poor stars (see
Crawford \& Barnes, 1970) all calibrations for metal-deficient stars
are extrapolations from the original standard system. Although much
effort is made to achieve agreement with the old standard system one
should be aware that these are extrapolations and that they might
differ because of differences in observational and data reduction
techniques used by different authors.

The two Str\"omgren systems for metal-deficient giants that are
commonly used are based on the catalogues by Olsen (1993) and Bond
(1980). The Bond catalogue was later extended by Anthony-Twarog \&
Twarog (1994).

A comparison between these two systems was published by Olsen (1995),
showing some systematic differences. For the $(b-y)$ and $m_{1}$
indices there is only a slight dependence on $(b-y)$ while the $c_{1}$
index shows a significant systematic difference on the order of 0.05
mag at $(b-y)$ = 0.04 and 0.02 mag at $(b-y)$ = 1.

Since we use secondary standard stars from Schuster \& Nissen (1988),
who used the Olsen (1993) standards to reduce their observations to
the standard system, our data will be tied to the Olsen system and
attention must be taken when comparing our observations with
observations or models based on any other system.

\begin{table}
      \caption[]{List of the photometry used for the 
 standard stars (from Schuster \& Nissen
	1988). Column 1 gives the star ID; Col. 2--5 give the
	magnitudes and colours. The full table is available
	electronically.}
         \label{tab.standards}
         \begin{tabular}{lllll}
            \hline
            \noalign{\smallskip}
            ID & $y_{0}$ & $(b - y)_{0}$ & $m_{1,0}$  & $c_{1,0}$  \\
            \noalign{\smallskip}
            \hline
            \noalign{\smallskip}
HD 33449     & 8.488  &   0.423  &   0.201  &   0.273 \\  
HD 46341     & 8.616  &   0.366  &   0.145  &   0.248 \\  
HD 51754     & 9.000  &   0.375  &   0.144  &   0.290 \\  

             \noalign{\smallskip}
            \hline
         \end{tabular}
   \end{table}
  
\section{Data reduction}
\label{sect:datared}
The entire dataset was processed using the INT Wide Field Survey
pipeline provided by the Cambridge Astronomical Survey Unit (Irwin \&
Lewis 2001). In addition to the usual calibrations and routines that
remove instrument signatures such as de-biasing, flat-fielding (using
dawn and dusk sky flats obtained each night during the observing run),
non-linearity, and gain corrections, the pipeline also provides tools
for photometric and astrometric calibrations as well as an object
catalogue. The astrometric solution is based on the Guide Star Catalog
and the accuracy is $\sim$ 1\arcsec.

The object detection was done using the described pipeline. Each
object that is detected is flagged as a star, an extended source, or
noise. In the following we will only consider objects that were
flagged as stars.

\subsection{Standard star photometry and transformations}
\label{sect:stand}

\begin{figure}
   \resizebox{\hsize}{!}{\includegraphics{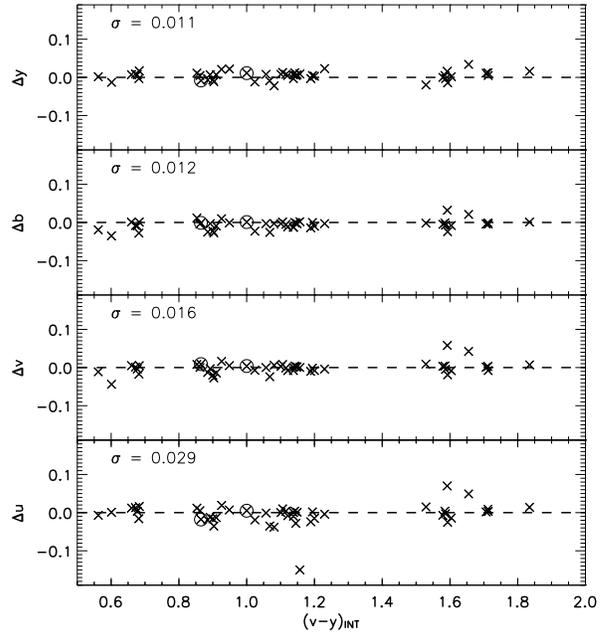}}   
\caption{The residuals for the standard stars as a function
of our final $(v-y)$, $\times$ (see Sect\,\ref{sect:stand}). The two
extinction stars are also included (marked by an extra $\circ$). 
For the extinction stars we have used the mean colours and magnitudes
based on all observations of these stars. }
\label{fig.res}%
\end{figure}

To obtain instrumental magnitudes for the standard stars we made
aperture photometry, using the IRAF\footnote{IRAF is distributed by
National Optical Astronomy Observatories, operated by the Association
of Universities for Research in Astronomy, Inc., under contract with
the National Science Foundation, USA.} {\sc daophot} package, on each
star in a small aperture, typically around 5 pixels in radius. Using a
curve-of-growth analysis we then corrected the magnitude to the radius
where the curve-of-growth converged.
  
The transformations from the instrumental system to the standard
Str\"omgren system were obtained by solving for the individual
magnitudes rather than the colour indices. This has the advantage that
we do not need to worry about the fact that observations through the
different filters are, for each standard star, obtained at
different airmasses (since we observe each filter separately, in
contrast to four-channel photometry where observations for all four
filters are obtained at the same time).  

First, we derived preliminary extinction coefficients, $k_i$, and zero
points, $z_i$, for each night in each filter, $i$, by solving the
following set of equations using the IRAF {\sc fitparam} task with a 2
$\sigma$ rejection after each fitting iteration:

\begin{eqnarray}
  i_{\rm s} & = & i_{\rm o}-k_{i}\cdot X - z_{i}
\end{eqnarray}

\noindent The standard magnitudes are on the left hand side (subscript
s), instrumental magnitudes on the right hand side (subscript o), $i$
denotes any of the four filters, and $X$ is the airmass.

We then applied the preliminary $z_i$ and $k_i$ to all our standard
stars from all nights and calculated preliminary colour terms, $a_i$,


\begin{eqnarray} 
i_{\rm s} & = & i_{\rm o}-a_{\rm i}\cdot (v-y)_{\rm s} - z_{\rm i}' \label{eq.a}
\end{eqnarray} 

\noindent where symbols are as in Eq.\,(1) and $a_{\rm i}$ are the
 colour coefficients for filter $i$. An additional zeropoint, $ z_{\rm
 i}'$, is introduced to improve the solution.

We then calculated a new set of $z_i$ and $k_i$ for each night using
 the preliminary colour terms derived above,
 
\begin{eqnarray}
i_{\rm s} & = & i_{\rm o}-a_{\rm i}\cdot (v-y)_{\rm s}-k_{\rm i}\cdot X - z_{\rm i} \label{eq.v}
\end{eqnarray} 
 
\noindent and then Eq.(\ref{eq.a}) and (\ref{eq.v}) are iterated in
this way until the solutions converged.  The final $z_i$, $k_i$ (for
each night), and $a_i$ are given in Table \ref{tab.ext}.

The residuals between our photometry and the standard values are shown
in Fig. \ref{fig.res}.  We note that the residuals are all small and
that the scatter is $\sim$ 0.01 for $v$, $b$, and $y$ and $\sim$ 0.03
magnitudes for $u$. No offsets are found nor any residual trends
with. The latter means that there is no need to add second-order terms
(compare with e.g. Fig.\,4 in Grundahl et al. (2002)).

\begin{table*}
\caption[]{Final extinction coefficients ($k_i$), zero-points ($z_i$), and colour term coefficients, ($a_{i}$)
for all filters as used in Eq.\,(\ref{eq.v}). Column 1 gives the dates when the observations were
done; Columns 2--13 give the extinction coefficients ($k_{i}$),
zero-points ($z_{i}$), and colour term coefficients, $a_{i}$ for each Str\"omgren filter $i$.}
\label{tab.ext}
\begin{tabular}{lllllllllllll}
\hline\hline      
Night & $ k_{y}$ & $z_{y}$ & $a_y$ &  $k_{b}$ &$z_{b}$ & $a_b $ & $ k_{v}$ & $z_{v}$ & $a_v$ & $k_{u}$ & $z_{u}$ & $a_u$  \\
\noalign{\smallskip}
\hline
\noalign{\smallskip}
  22 & 0.114 & --22.768 & 0.002  & 0.176 &  --23.207& 0.020  & 0.275 & --23.149 & 0.045  & 0.521 & --23.106 & 0.077 \\
  23 & 0.124 & --22.784 & 0.002  & 0.180 &  --23.210& 0.020  & 0.291 & --23.144 & 0.045  & 0.546 & --23.103 & 0.077 \\
  24 & 0.111 & --22.749 & 0.002  & 0.174 &  --23.191& 0.020  & 0.284 & --23.127 & 0.045  & 0.528 & --23.070 & 0.077 \\
  25 & 0.117 & --22.782 & 0.002  & 0.173 &  --23.218& 0.020  & 0.295 & --23.162 & 0.045  & 0.529 & --23.092 & 0.077 \\
  28 & 0.108 & --22.745 & 0.002  & 0.180 &  --23.201& 0.020  & 0.284 & --23.118 & 0.045  & 0.502 & --22.986 & 0.077 \\
\noalign{\smallskip}
\hline
\end{tabular}
\end{table*}

\subsection{Photometry for Draco}
\label{sect:photanderrors}

\begin{figure}
\resizebox{\hsize}{!}{\includegraphics{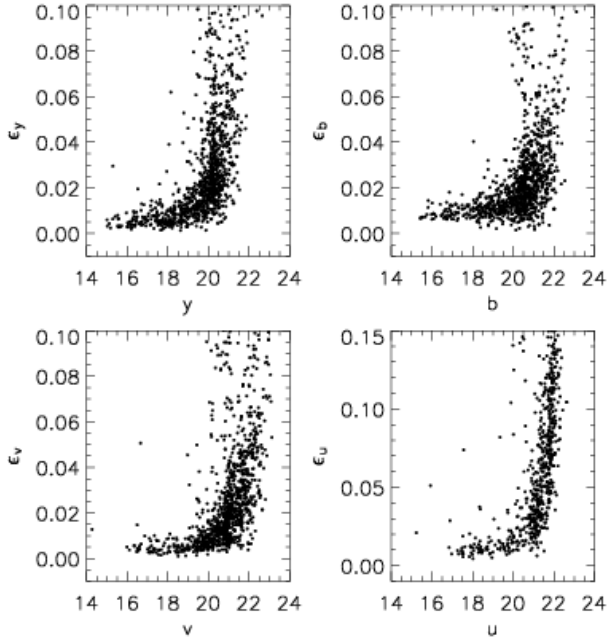}}
\caption{The resulting errors in our magnitudes 
for the four filters. Note the different scale
on the vertical axis for $u$. }
\label{fig.errors}%
\end{figure}

\begin{figure}[t!]
\resizebox{\hsize}{!}{\includegraphics{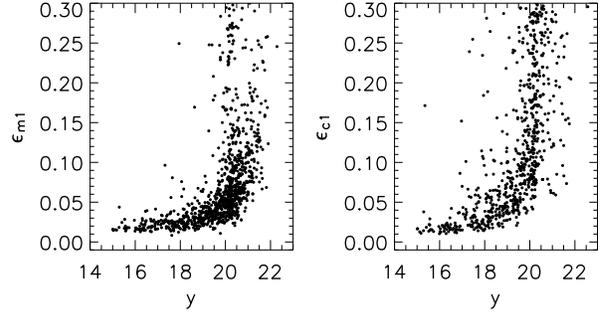}}
\caption{Errors in the $m_{\rm 1}$ and $c_{\rm 1}$ indices as a
function of the $y$ magnitude.}
\label{fig.m1c1err}
\end{figure}

For the science frames of the Draco dSph galaxy we again obtained
aperture photometry using the {\sc daophot} package in IRAF. An
aperture radius of 5 pixels (1\farcs65) was used to minimize any
effects of crowding. Using a curve of growth derived from a number of
bright isolated stars on each frame we then corrected the measured
magnitudes out to where the curve of growth converged. After
correcting for airmass and applying zero-points for each night the
photometry from the individual images were merged by averaging.

To avoid errors arising from the fact that we made our photometry on
the individual frames rather than on a combined, cosmic ray cleaned
frame we used the following iterative method. A mean magnitude was
calculated and individual measurements falling outside $\pm 3\sigma$
of this mean were rejected. The mean was recalculated and rejections
were made again.  For $y$, $b$, and $v$ the second rejection was at
the $\pm 1.25\sigma$-level while for $u$ we applied a second rejection level
of $\pm 1.5\sigma$. The magnitudes were then converted to the standard
Str\"omgren system using Eq.\,(\ref{eq.v}) with the coefficients
listed in Table\,\ref{tab.ext}. 

The photometric error for star $i$ was calculated as the error in the
mean, which is defined as $\epsilon_{i} = \sigma_{i}/ \sqrt{n_{i}}$,
where $n_{i}$ is the number of measurements kept after the rejection
process, and $\sigma_{i}$ is the standard deviation for those measurements. 

The resulting errors are shown in
Fig.\,\ref{fig.errors}.  Note that the error in $u$ is larger than in
the other filters since those observations are not as deep as the
others. In Fig.\,\ref{fig.m1c1err} we show the resulting errors on the
$m_{1} = (v-b) - (b-y)$ and $c_{1} = (u-v) - (v-b)$ Str\"omgren
indices.

\begin{figure*}
\resizebox{\hsize}{!}{\includegraphics{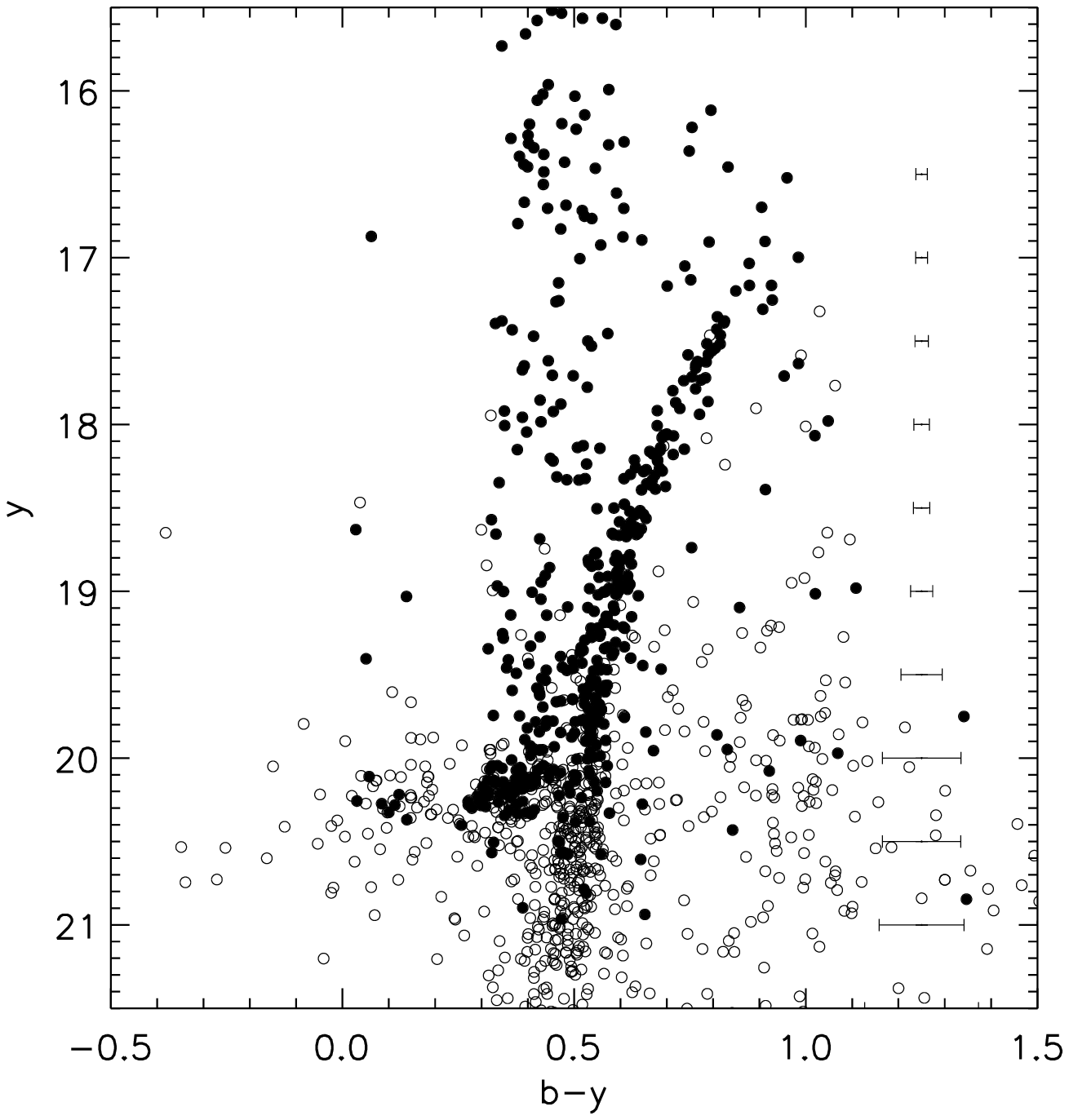}\includegraphics{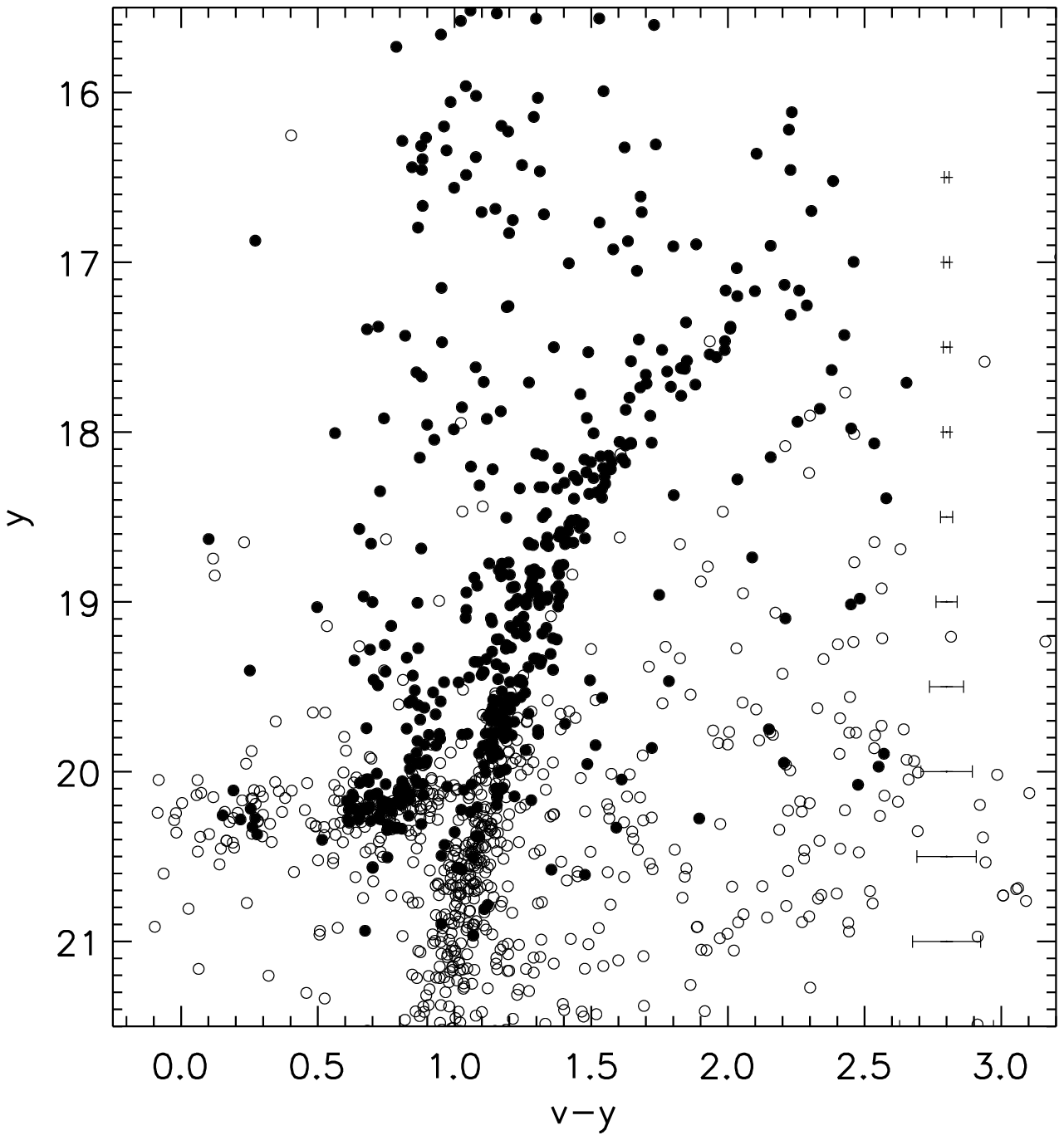}}   
\caption{$y$ vs $(b-y)$ and $y$ vs $(v-y)$ colour-magnitude diagrams
for the Draco dSph galaxy. $\circ$ mark all stars that were measured
regardless of errors and $\bullet $ mark the stars that have $y<21$,
$\epsilon_{(b-y)}<0.18$, and $\epsilon_{c_{\rm 1}}<0.18$ (see
Sect.\,\ref{sect:select.stars}). Error bars to the right indicate
typical errors in the colour at that magnitude.}
\label{fig.cmds}%
\end{figure*}

\subsection{Comparison with photometry from other studies}

\begin{figure}
\resizebox{\hsize}{!}{\includegraphics{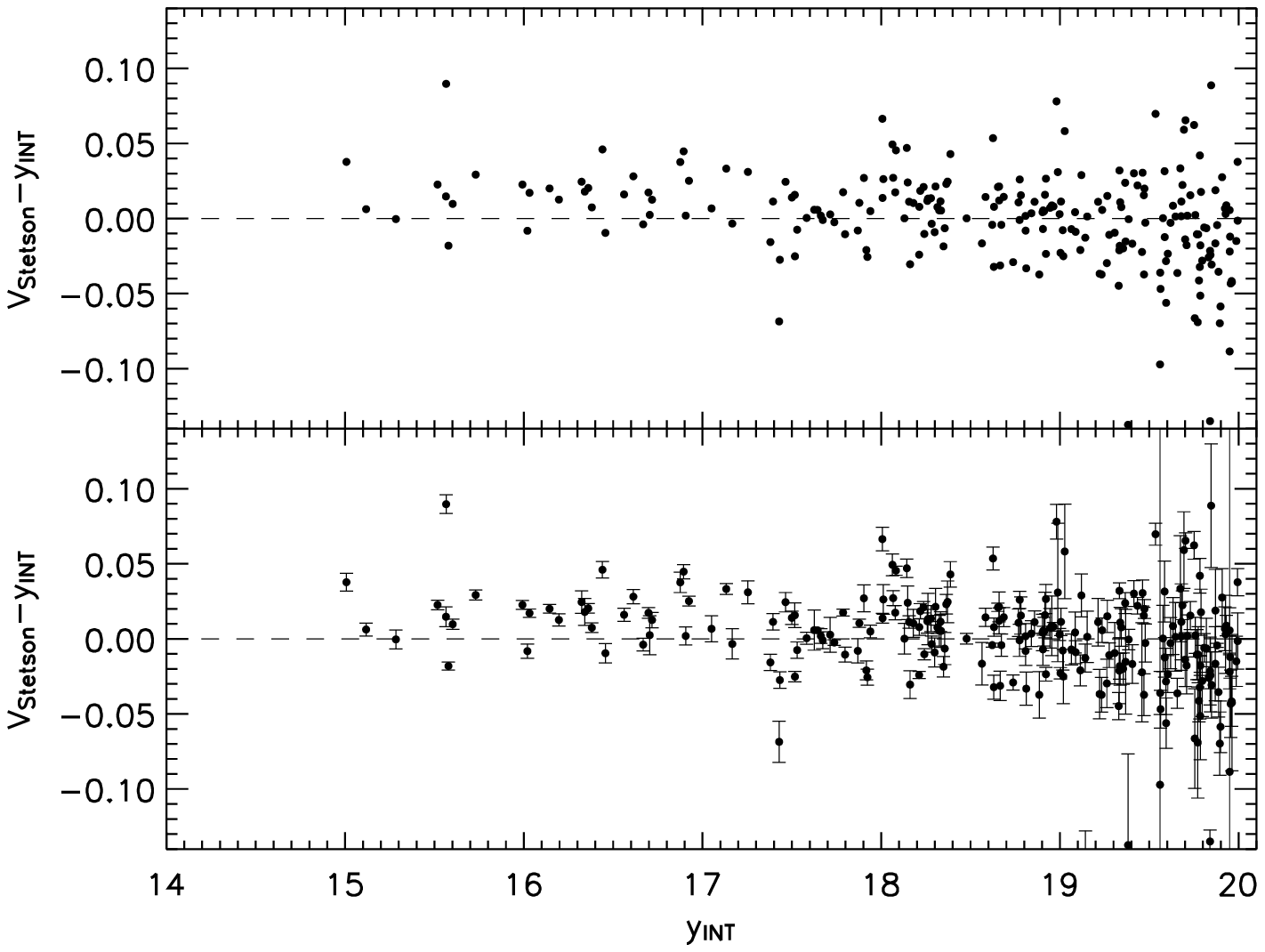}}
\caption{A comparison of our $V$ ($y_{\rm INT}$) magnitudes and $V$
magnitudes  by P. Stetson ($V_{\rm Stetson}$) for stars in our field centered
on the Draco  dSph galaxy
as a function of $y_{\rm INT}$. 
For stars with $15< y_{\rm INT} < 17$ the offset is 
0.018 magnitudes with a scatter of 0.021, while for stars with 
$17< y_{\rm INT} < 19$ the offset is 
0.007 magnitudes with a scatter of 0.021.
The lower panel also includes the error-bars, which were left out
from the upper panel for clarity.
The broad
band photometry is available at {\tt http://cadcwww.hia.nrc.ca/standards/}. }
\label{fig.comp.y.VS}%
\end{figure}

\begin{figure}
\resizebox{\hsize}{!}{\includegraphics{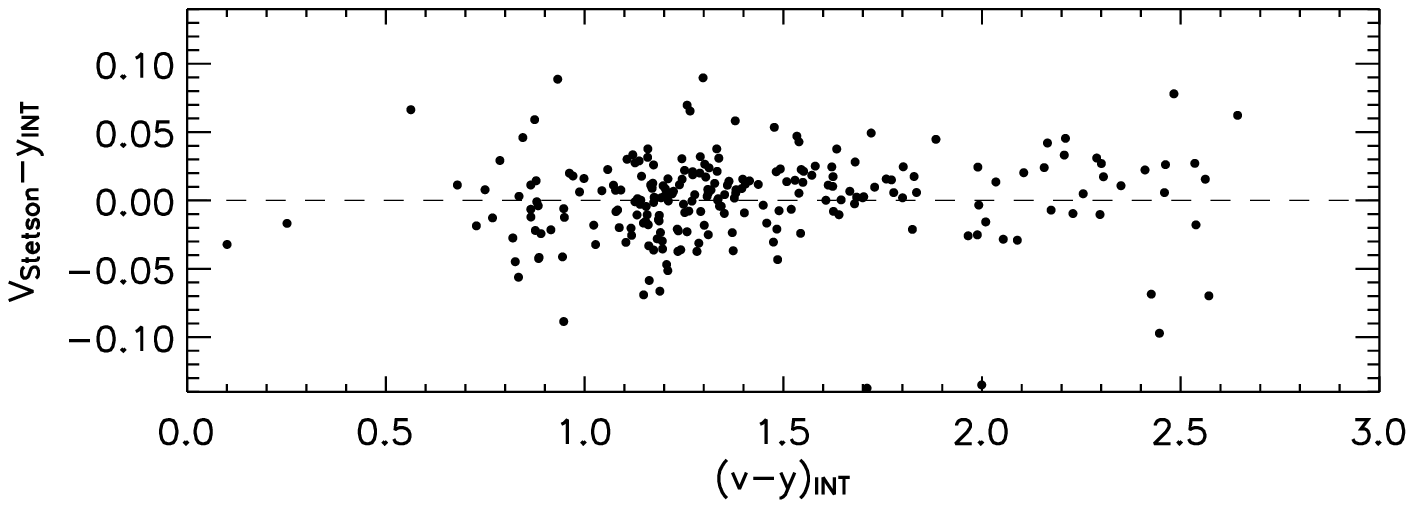}}
\caption{A comparison of our $V$ ($y_{\rm INT}$) magnitudes and $V$
magnitudes  by P. Stetson ($V_{\rm Stetson}$) for for stars in our field
centred on the Draco dSph galaxy
as a function of $(v-y)_{\rm INT}$. 
The broad
band photometry is available at {\tt http://cadcwww.hia.nrc.ca/standards/} }
\label{fig.comp.y.VS.vy}%
\end{figure}

As an independent assessment of the quality of our data we have made a
comparison between our $y$ (where we assume Str\"omgren $y$ = V; Olsen
1983) and the V magnitudes derived by P. Stetson.
The photometric data from Stetson are available at {\tt
  http://cadcwww.hia.nrc.ca/standards/}. Figures \ref{fig.comp.y.VS}
and \ref{fig.comp.y.VS.vy} show comparisons between our $y$ magnitudes
and the Stetson $V$ magnitudes for stars in our field centred on the
Draco dSph galaxy. We only make the comparison for stars that are
brighter than $V=20.0$.

The differences between our and Stetson's magnitudes appear to be 
an only an offset without any colour dependence. The offset is 
0.018 magnitudes when the brighter stars are considered. We take this 
as a measure of the absolute error in the calibration of our
 $y$ magnitudes. That the magnitude difference decreases as the 
magnitudes increase reflects the fact that the 
two studies reach different depths with the same accuracy.

\subsection{The colour-magnitude diagram}
Figure \ref{fig.cmds} presents our $y$ vs $(b-y)$ and $y$ vs $(v-y)$
colour magnitude diagrams (CMD) for the stars in our field centred on
the Draco dSph galaxy.  The most prominent feature is the well defined
RGB from $y=17$ down to $y \simeq 21.5$. A well populated horizontal
branch (HB) is seen at $y \simeq 20.2$. Note the gap along the HB at
$(b-y) \simeq 0.2-0.3$ and $(v-y) \simeq 0.4-0.5$ caused by the random
colour and magnitude variations of the RR Lyrae stars populating this
region.

A fair number of foreground objects can also be seen with $(b-y)$ in
the range expected for foreground dwarf stars.  These stars should
mainly be situated in the thick disk and the halo of the Milky
Way. The CMD shows the typical sharp cut-off at $(b-y) \simeq 0.3$ and
$(v-y) \simeq 0.65$ associated with the blue limit of the turnoff
stars. The various stellar populations present in the diagrams in
Fig.\,\ref{fig.cmds} will be further discussed in
Sect.\,\ref{sect:whichpops}.

\subsection{Reddening}
\label{sect:red}

The reddening towards the Draco dSph galaxy is small. Stetson (1979)
performed a detailed study of the reddening of stars along the line of
sight towards the Draco dSph galaxy. He concluded that the total
reddening (which, as the Draco dSph galaxy is essentially dust free,
must emanate from the Milky Way) is $E(B-V)=0.03\pm 0.01$. We adopt
this value in this paper.

The reddenings for the Str\"omgren indices are taken from Schlegel et
al. (1998). Their scale is based on the commonly used $R_{\rm V}=3.1$
from Cardelli et al. (1989).  Using the Schlegel et al. (1998)
relations a reddening of $E(B-V)=0.03$ corresponds to $E(b-y)=0.023$,
$E(v-y)=0.038$, $E(m_{\rm 1})=-0.008$, and $E(c_{\rm 1})=0.005$. 

\section{Identifying Draco members and categorising the field stars}
\label{sect:identifying}

Although the Draco dSph galaxy is at a high latitude and hence the
foreground contamination is not as severe as it is in other lines of
sight it is still not negligible.  It is our aim to use the RGB to
study the properties of the stellar populations in the Draco dSph
galaxy. Therefore it is very important to confirm that the stars that
appear to be on the RGB in the Draco dSph galaxy indeed are members of
the dSph.

\begin{figure}[t!]
\resizebox{\hsize}{!}{\includegraphics{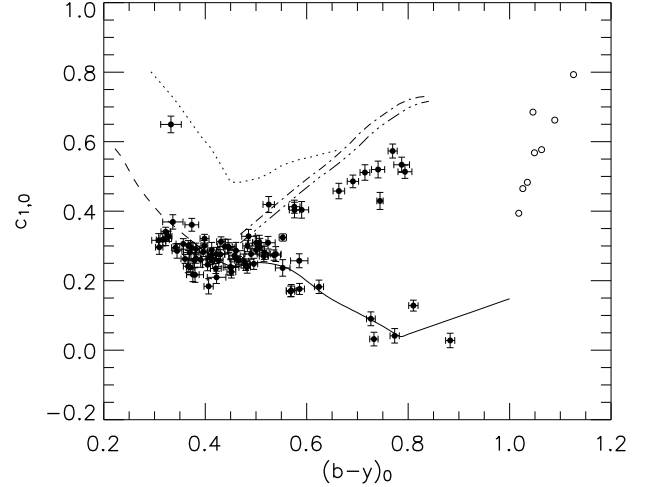}}
\caption{$c_{\rm 1,0}$ vs $(b-y)_0$ diagram for stars with $y<21$ and
$\epsilon_{c_{\rm 1}} < 0.025$.  Errors as indicated.  The data have
been corrected for reddening according to Sect. \ref{sect:red}.  The
standard relations for dwarf stars from Crawford (1975) and Olsen
(1984) are shown with a dashed and full line, respectively. The dotted
line indicates the RHB/AGB according to Anthony-Twarog \& Twarog
(1994) and the long-dash-short-dash and the long-dash-triple-dot lines
indicate the RGB according to Anthony-Twarog \& Twarog (1994) for
[Fe/H] = --2 and --1.5 dex, respectively. $\circ$  denotes the stars
used by Olsen (1984) to trace the MIII giants. The relations from
Anthony-Twarog \& Twarog (1994) have been corrected to the same system
as Olsen (1993) using Olsen (1995). See also discussion in Sect. \ref{sect:photsystem}.}
\label{fig.c1.by.errc10.05}
\end{figure}

\begin{figure}[t!]
\resizebox{\hsize}{!}{\includegraphics{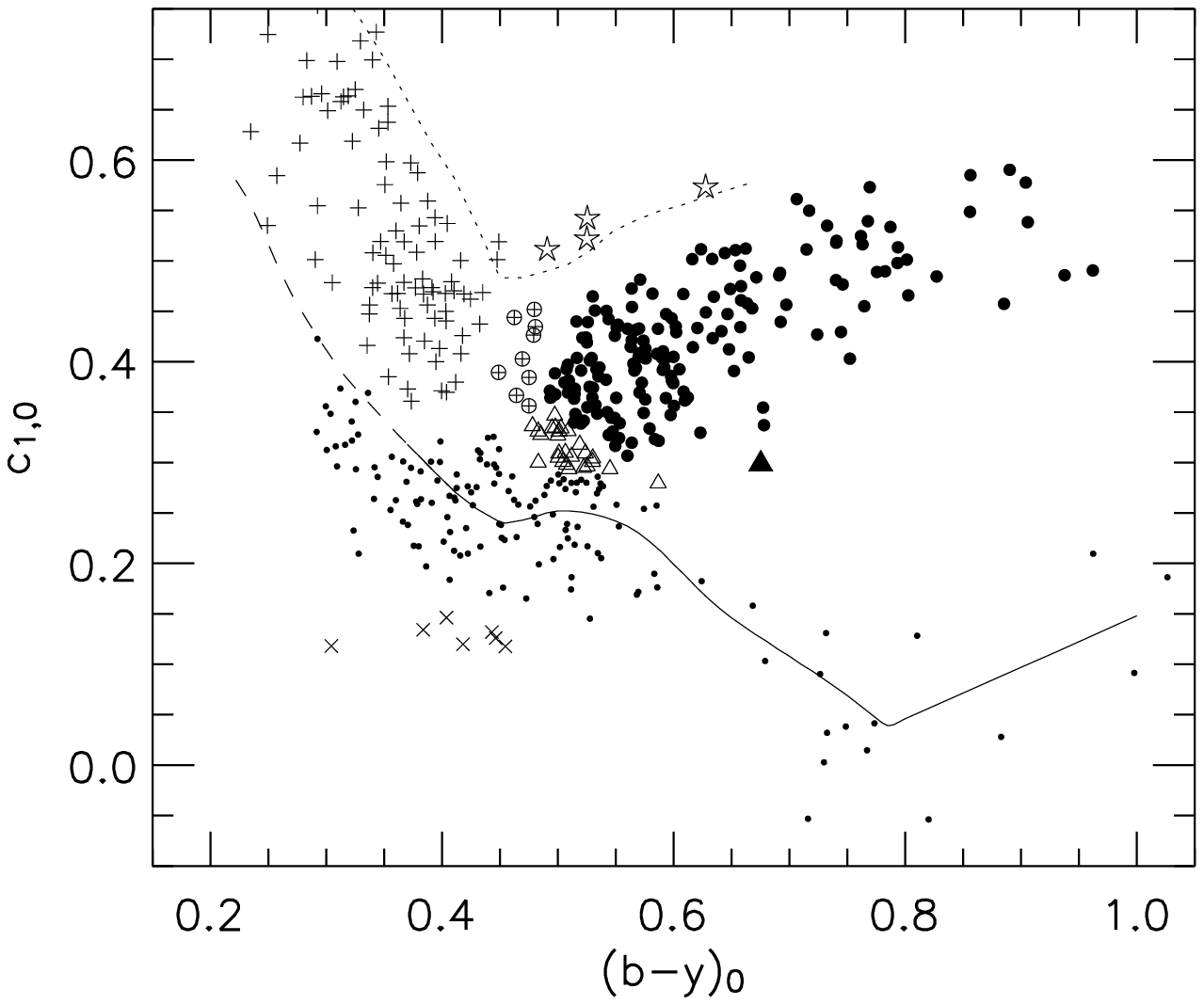}}
\caption{$c_{\rm 1,0}$ vs $(b-y)_0$ diagram for stars with $y<21$,
$\epsilon_{(b-y)}<0.18$, and $\epsilon_{c_{\rm 1}}<0.18$.  The data
have been corrected for reddening according to Sect. \ref{sect:red}.
The standard relations for dwarf stars from Crawford (1975) and Olsen
(1984) are shown with a dashed and full line, respectively. The dotted
line indicates the RHB/AGB according to Anthony-Twarog \& Twarog
(1994). The relations from Anthony-Twarog \& Twarog (1994) have been
corrected to the same system as Olsen (1993) using Olsen (1995). See
also discussion in Sect. \ref{sect:photsystem}. Large $\bullet$
indicate RGB stars, small $\bullet$  dwarf stars, $\blacktriangle$
indicates the one known carbon star inside the limits of the plot,
$\times$ a small group of likely foreground dwarf stars
falling below the dwarf sequence, $\oplus$ stars that cannot be easily
assigned as HB or RGB stars, $\vartriangle$ stars that cannot be
easily assigned as dwarf or RGB stars, open stars indicate possible
AGB stars, and, finally, $+$ designate HB stars.}
\label{fig.c1.by.zoom.marked}
\end{figure}

\begin{figure}[t!]
\resizebox{\hsize}{!}{\includegraphics{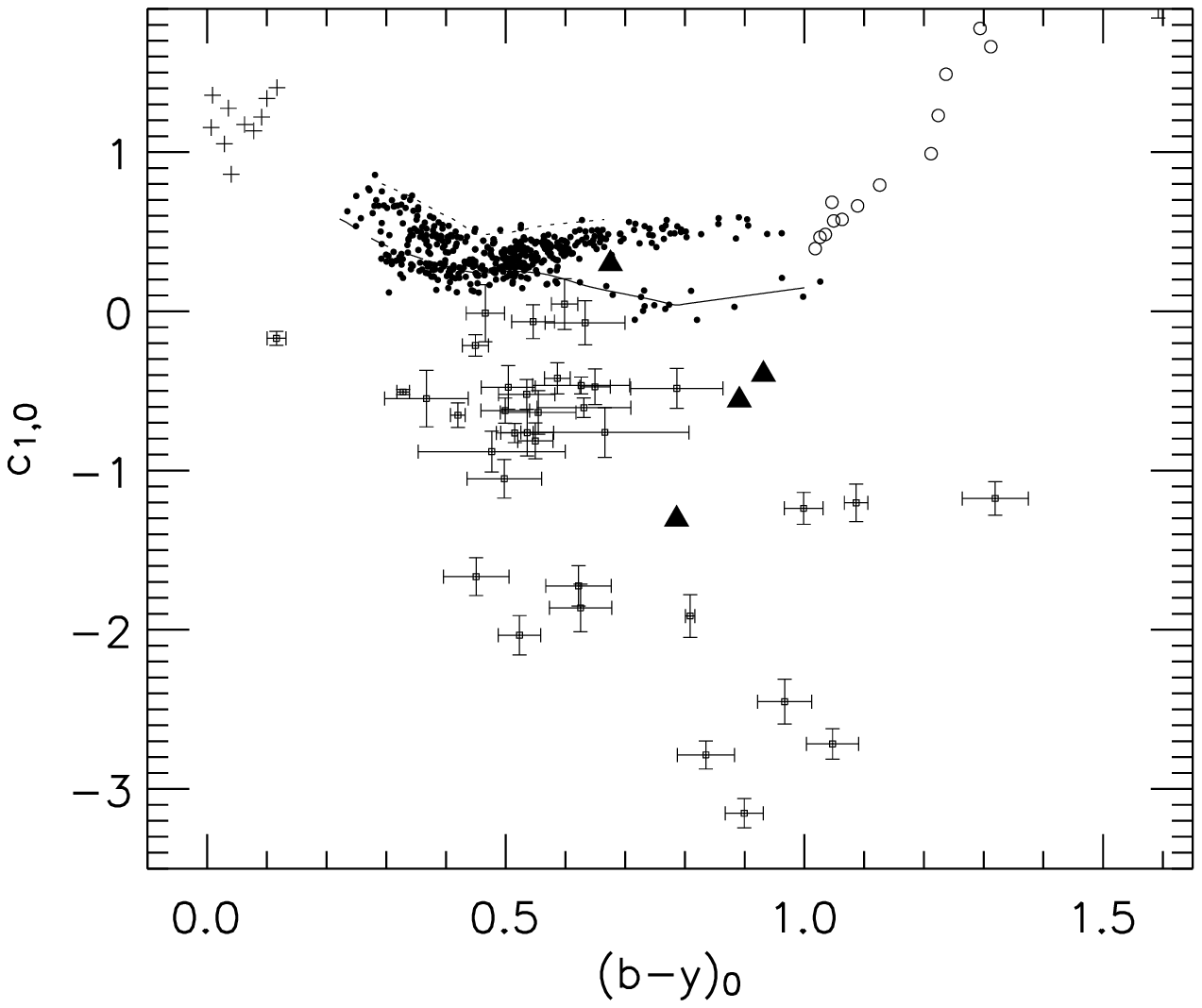}}
\caption{Full view of the $c_{\rm 1,0}$ vs $(b-y)_0$ plane for stars
 with $y<21$, $\epsilon_{(b-y)}<0.18$, and $\epsilon_{c_{\rm
 1}}<0.18$.  The data have been corrected for reddening according to
 Sect. \ref{sect:red}.  The standard relations for dwarf stars from
 Crawford (1975) and Olsen (1984) are shown with a full and dashed
 line respectively. The dotted line indicates the RHB/AGB according to
 Anthony-Twarog \& Twarog (1994).  The relations from Anthony-Twarog
 \& Twarog (1994) have been corrected to the same system as Olsen
 (1993) using Olsen (1995).  Stars that are neither on the dwarf or
 giant branches are displayed with error-bars. $\blacktriangle$ marks
 the known carbon stars, see Table \ref{tab.carbon}, $+$ indicate
 probable blue HB stars or hot main sequence stars. $\circ$  denote
 the stars used by Olsen (1984) to trace the MIII giants.}
\label{fig.c1.by.fullview}
\end{figure}

The Str\"omgren photometric system gives us opportunities to identify
stars at different evolutionary stages without knowing the distances
to them.  The $c_{\rm 1}$ index in the Str\"omgren system is defined
to measure the Balmer discontinuity in a stellar spectrum. The
blocking in the Str\"omgren $u$-band from metal lines is approximately
twice that in the $v$-band (for a schematic diagram of this see
e.g. Golay 1974, Fig. 116). Since $c_{\rm 1}=(u-v)-(v-b)$, this
difference is taken care of in the construction of the index (with
some remaining dependence on metallicity, see e.g. Gustafsson \& Bell
1979, Fig. 23).

\subsection{Selection of stars for further study}
\label{sect:select.stars}
We now want to make a selection of the stars that we will base our
discussions on. The three filters $v$, $b$, and $y$ all have
significantly smaller errors than $u$. As we want to disentangle the
different stellar populations present in the colour-magnitude diagram
using the $c_{\rm 1}$ index we will be governed by the error in $u$
when selecting stars for further analysis. We later compare the
results from the $c_{\rm 1}$ investigation with results from studies
of radial velocity and proper motion membership. This will provide a
measure of how successful our approach is and if the somewhat
shallower $u$ observations are a limiting factor in our study.  In
Fig.\,\ref{fig.cmds} all stars with $y<21$, $\epsilon_{(b-y)}<0.18$,
and $\epsilon_{c_{\rm 1}}<0.18$ are indicated as filled circles. These
are the stars that we will consider in the following sections.

\subsection{Which stellar populations do we see in the colour-magnitude diagram?}
\label{sect:whichpops}

\subsubsection{Finding the different populations}
\label{sect:c1}

We start by considering the stars for which we have the very best
photometry.  In Fig.\,\ref{fig.c1.by.errc10.05} we show the c$_{\rm
1}$ vs $(b-y)$ diagram for stars with $y<21$ and
$\epsilon_{c_{\rm 1}} < 0.025$.  

Crawford (1975) and Olsen (1984) provide (preliminary) standard
relations for field dwarf stars and a large number of stars in our
data set follow these relations nicely (see
Fig.\,\ref{fig.c1.by.errc10.05}).

The relations for RGB and red horizontal branch (RHB) and asymptotic
giant branch (AGB) stars are taken from Anthony-Twarog \& Twarog
(1994) and have been corrected to the system of Olsen (1993) using the
relations in Olsen (1995), see Sect. \ref{sect:photsystem}. We find
that no star falls on the AGB sequence and that there is one probable
RHB star at $(b-y)_{\rm 0} \sim 0.33$ and $c_{\rm 1,0} \sim 0.65$.  A
number of RGB stars fall just under the relations from Anthony-Twarog
\& Twarog (1994).

As giant stars with lower metallicities, at a given $(b-y)$, have
higher $c_{\rm 1}$ (compare the relations shown in the figure and,
more importantly, the results in Gustafsson \& Bell 1979) this would
indicate that the Draco dSph galaxy was significantly more metal-rich
than [Fe/H]$=-1.5$. This appears unlikely as investigations of the
metallicity, in a limited number of the RGB stars in the Draco dSph
galaxy, have shown the metallicity of the dSph to be between $-2$ and
$-1.5$ (e.g. Shetrone et al. 2001a; Bell 1985; Zinn 1978). A more
likely explanation is that the empirical relations overestimate the
$c_{\rm 1}$ index for RGB stars at a given metallicity.

Also shown are the stars used by Olsen (1984) to define the relation
for MIII stars. As can be seen none of our stars (with very good
photometry) have such red colours.

Three stars with $y<21$ and $\epsilon_{c_{\rm 1}} < 0.025$ fall outside
the boundaries of Fig.\,\ref{fig.c1.by.errc10.05}. These are (\# INT,
$(b-y)_0$, ${c_{\rm 1,0}}$, $\epsilon_{\rm (b-y)}$, $\epsilon_{c_{\rm 1}}$)=
(\#240, 0.328, --0.507, 0.010, 0.016), (\#1961, 0.007, 1.154, 0.015,
0.020), (\#1984, 0.040, 0.860, 0.012, 0.016). Stars \# 1961 and 1984
fall on the continuation of the ZAMS and the HB sequence and are hence
hotter foreground stars or blue HB stars. Star \# 240 appears peculiar
in its ${c_{\rm 1}}$, we have no simple explanation for this.

After having demonstrated how the relations in the c$_{1}$ vs $(b-y)$
plane work and where our stars with the very best photometry fall we
turn to an investigation of the larger sample defined by $y<21$,
$\epsilon_{(b-y)}<0.18$, and $\epsilon_{c_{\rm 1}}<0.18$ as shown in
Fig.\,\ref{fig.c1.by.zoom.marked}. This plot shows, as expected,
larger scatter than Fig.\,\ref{fig.c1.by.errc10.05}. Nevertheless, the
various standard relations are still well traced and we can easily
identify the foreground dwarf stars, the RGB and RHB stars in the
Draco dSph galaxy.  Figure \,\ref{fig.c1.by.fullview} is equivalent to
Fig.\,\ref{fig.c1.by.zoom.marked} but shows the full range in ${c_{\rm
    1,0}}$ and $(b-y)_{\rm 0}$ covered by our complete sample.

\paragraph{Foreground dwarf stars}

For dwarf stars there is some metallicity dependence in the $c_{\rm
  1,0}$ vs $(b-y)_{0}$ relation such that starting from around
$(b-y)_0$ = 0.25 the less metal-rich stars fall below the solar
metallicty ZAMS. This is illustrated in e.g. Fig. 13 and 14 in Clem et
al. (2004). Hence we have also selected stars that fall below as well
as above the relation from Olsen (1984) to be dwarf stars. Compare
also our Fig.\,\ref{fig.c1.by.errc10.05}, \ref{fig.c1.by.zoom.marked}
and \ref{fig.c1.by.fullview} with Fig. 8 in Olsen (1984).

\begin{figure*}[t!]
\resizebox{\hsize}{!}{\includegraphics{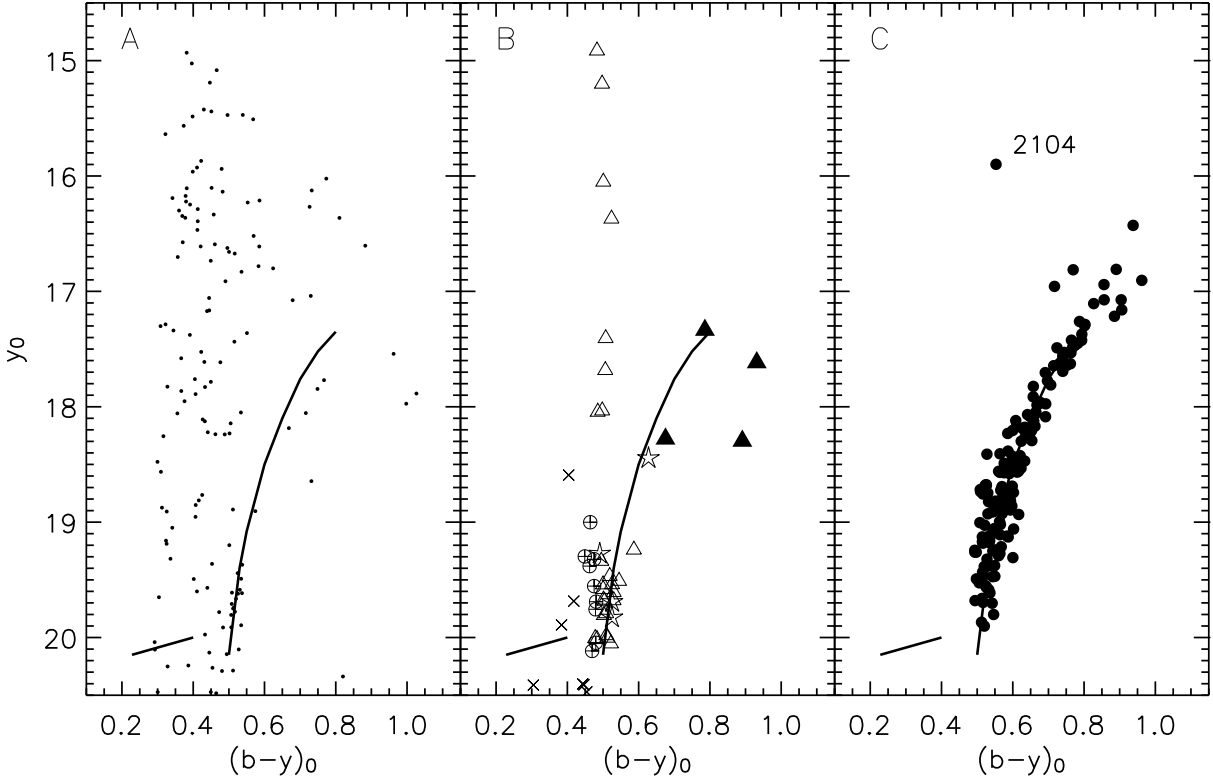}}
\caption{$y_0$ vs $(b-y)_0$ colour-magnitude diagrams for stars with
$y<21$, $\epsilon_{(b-y)}<0.18$, and $\epsilon_{(b-y)}<0.18$. \textbf{A)}
shows stars identified as dwarf stars. \textbf{B)} $\triangle$ indicates stars
that are not easily attributed to either the dwarf sequence or the
RGB, $\times$ indicate a group of likely foreground dwarf
stars,$\oplus$ stars that cannot be easily
assigned as HB or RGB stars, open stars indicates possible AGB stars, and $\blacktriangle$
indicate known carbon stars (see Table\,\ref{tab.carbon}). \textbf{C)} Stars
identified as RGB stars. The solid lines show tracings of the RGB and
HB, and are included to guide the eye.}
\label{fig.cmd.dwarfs.giants}
\end{figure*}

Figure \ref{fig.cmd.dwarfs.giants}A shows the resulting CMD
for stars classified as dwarf stars.

\paragraph{Members of the RGB in the Draco dSph galaxy} 
As discussed earlier the RGB stars can be distinguished from the dwarf
stars with the same colour with the help of the $c_{\rm 1}$ index. In
Fig.\,\ref{fig.c1.by.zoom.marked} we have identified RGB stars in the
Draco dSph galaxy in this way.

Around $(b-y)_0 \sim 0.5$ there is significant uncertainty in
evolutionary status for values of $c_{\rm 1,0}$ around 0.3. The same
is true for stars with $(b-y)_0 \sim 0.45$ and $c_{\rm 1,0} \sim 0.4$
where it is not clear if they belong to the RGB or HB sequence. We
have handled this by collecting those stars into separate groups as
indicated in Fig. \,\ref{fig.c1.by.zoom.marked}. Figure
\ref{fig.cmd.dwarfs.giants}B shows where these stars fall in the CMD.
That our cautionary treatment of these stars is justified is
exemplified by a clustering of stars around the RGB in the Draco dSph
galaxy as well as a scattering towards brighter $y$ magnitudes.  For
the RGB stars in the Draco dSph galaxy this gives a first indication
of the $(b-y)$ level at which the usage of the $c_{\rm 1} $ index as a
luminosity indicator breaks down. In this case at $(b-y)_{0}$ $\sim$
0.5 (compare Fig.\,\ref{fig.cmd.dwarfs.giants}B).

Finally, Fig. \ref{fig.cmd.dwarfs.giants}C shows the
resulting CMD for the RGB in the Draco. One star, \#2104, is far brighter than
the expected RGB with $y_0 \sim 15.9$. On closer inspection of this
star's position in the $c_{\rm 1,0}$ vs $(b-y)_0$ diagram we find that it is 
located close to the stars we that have flagged as being of uncertain
status. It is thus possible that this
star is a mis-classified dwarf star. Another possibility is that this star
is a faint foreground giant or sub-giant star. Such stars have a typical
absolute magnitude M$_{V} \sim +2$, which means that this star would be
at a vertical height of $\sim$ 3.4 kpc above the galactic plane
(adopting a value of the galactic latitude in the direction of the Draco dSph galaxy,
$b$ = +34.7$^\circ$ from Mateo (1998)).

\paragraph{Probable AGB stars?}
Based on the relation from Anthony-Twarog \& Twarog (1994) we have
identified four stars as probable AGB stars. However, in the CMD in
Fig.\,\ref{fig.cmd.dwarfs.giants}B they do not behave
like AGB stars (i.e. located along an expected AGB sequence on the blue
side of the RGB). 

\paragraph{Carbon stars}
Five of the stars in the Draco dSph galaxy are known carbon stars
(Table\,\ref{tab.carbon}). Four of these stars have $y<21$,
$\epsilon_{(b-y)}<0.18$, and $\epsilon_{c_{\rm 1}}<0.18$ and
have been explicitly identified
in Figs.\,\ref{fig.c1.by.zoom.marked}, \ref{fig.c1.by.fullview}, and
\ref{fig.cmd.dwarfs.giants}. Three of them have low $c_{\rm 1,0}$
and could therefore not be confused with the RGB stars in the $c_{\rm
1,0}$ vs $(b-y)_0$ diagram. We note that there are a number of stars
in similar positions in the $c_{\rm 1,0}$ vs $(b-y)_0$ diagram and one
may speculate that some of them are carbon stars. This should be further
investigated using spectroscopy.

\paragraph{Variable stars}
Our data is single-epoch and do not give any information as to the
variability of the observed stars. Most of the known variables in the
Draco dSph galaxy were identified by Baade \& Swope (1961) using an
area smaller than that covered by our images. This means that outside
their image we have no knowledge about stellar variability.  Our data
were cross-correlated with that of Baade \& Swope (1961) so that all
known variables within our data were identified. None of our RGB
members are known variables.

\begin{table}
\caption[]{Known carbon stars in the Draco dSph galaxy. Column 1 gives our ID; Col. 2
the name in Aaronson et al. (1982); Col. 3 the reference to each
star.}
\label{tab.carbon}
\begin{tabular}{llll}
\hline\hline
\noalign{\smallskip}
INT ID & Name & Reference \\
\noalign{\smallskip}
\hline
\noalign{\smallskip}
1038 & C4  & Aaronson et al. (1982)\\
1119 & C3  & Aaronson et al. (1982)\\
2038 & C2  & Aaronson et al. (1982)\\
2094 & C1  & Aaronson et al. (1982)\\
2127 &     & Shetrone et al. (2001b)\\
\noalign{\smallskip}
\hline
\end{tabular}
\end{table}

\subsection{Other membership criteria}

\subsubsection{Radial velocities}
\label{sect:radvel}

\begin{figure}
\resizebox{\hsize}{!}{
\includegraphics{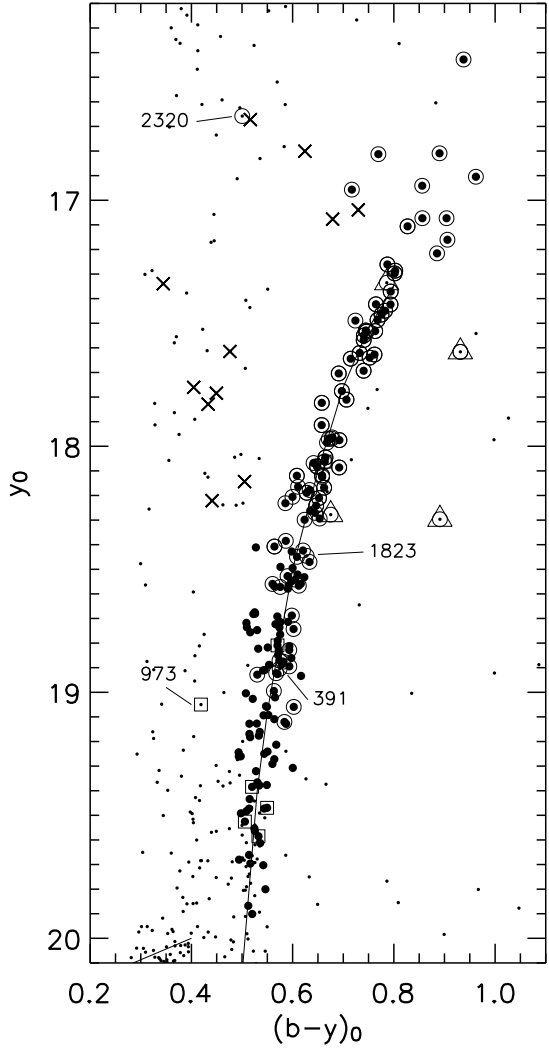}}
\caption{$y_0$ vs $(b-y)_0$ colour-magnitude diagram for the Draco dSph galaxy and
field stars cross-correlated with radial velocity studies. All stars
have $y<21$, $\epsilon_{(b-y)}<0.18$, and $\epsilon_{c_{\rm 1}}<0.18$.
$\bullet$  mark our RGB stars, $\circ$ mark radial velocity members
from Kleyna et al. (2002) and Armandroff et al. (1995), $\square$ mark
probable members from Kleyna et al., and $\times$ mark non-members
from Armandroff et al. (1995) ID numbers for stars discussed in the text are
included. Also shown are known carbon stars ($\triangle$). Note that
the axes scales are different from previous figures. The solid lines
show a fiducial of the RGB and HB to guide the eye.}
\label{fig.cmd.RVmembers}
\end{figure}

\begin{figure}
\resizebox{\hsize}{!}{\includegraphics{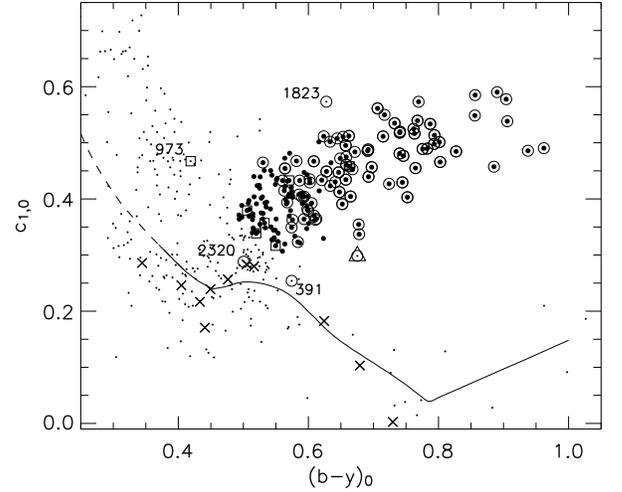}}
\caption{$c_{\rm 1,0}$ vs $(b-y)_0$ diagrams cross-correlated with
radial velocity studies. $\bullet$  mark our RGB stars, $\circ$ mark
radial velocity members from Kleyna et al. (2002) and Armandroff et
al. (1995), $\square$ mark probable members from Kleyna et al., and
$\times$ mark non-members from Armandroff et al. ID numbers of stars
discussed in the text are included. Also shown are known carbon stars
($\triangle$). The standard relations for dwarf stars from Crawford
(1975) and Olsen (1984) are shown with dashed and full lines,
respectively.}
\label{fig_c1.RVmembers}
\end{figure}

There are two studies in the literature with radial velocity
measurements for stars in the direction of the Draco dSph galaxy: Kleyna et
al. (2002) and Armandroff et al. (1995).  We can utilize this
information to test the procedure we used in the previous section to
isolate member stars.  

Since the Draco dSph galaxy (as do other dSphs) has a substantial
velocity relative to the Milky Way (Mateo 1998), radial velocities of
individual stars might appear as the ultimate tool to identify those
that are truly members of the dSph. However, radial velocities are of
limited value for (at least) two reasons: the presence of binary
systems and stellar activity (Armandroff et al. 1995). Also, while
radial velocity measurements can be efficiently used in the central
regions of a dSph, where a large majority of the stars along the
expected RGB are members, in the outer regions it is very time
consuming to identify only a few real members projected onto a large
amount of foreground contamination.

The most recent radial velocity study is by Kleyna et al. (2002)
who list 159 members with well measured radial velocities and 27
members with less well-determined velocities. In the earlier study by
Armandroff et al. (1995), an additional 91 radial velocity members are listed.
Because of partial overlap in the two studies, the total number of
unique radial velocity members available is 188.

Figures \, \ref{fig.cmd.RVmembers} and \, \ref{fig_c1.RVmembers} show
the CMD and $c_{\rm 1,0}$ vs $(b-y)_0$ diagrams for our sample of
members in the Draco dSph galaxy cross-correlated with the radial
velocity studies by Kleyna et al. (2002) and Armandroff et
al. (1995). Out of the 188 radial velocity members 88 lie within our
field and fulfill $y<21$, $\epsilon_{(b-y)}<0.18$, and
$\epsilon_{c_{\rm 1}}<0.18$. All but seven of these stars were
selected as RGB members when we used the $c_{\rm 1}$ index.

Four of the discrepant stars not in our RGB sample (\#1038, \#1119,
\#2094, and \#2127) are known carbon stars listed in Table 6 and hence
excluded from our RGB sample although they are likely members of the
Draco dSph galaxy.

Star \#391 falls on the RGB sequence in the CMD but is identified as a
dwarf because of its position on the dwarf star sequence in the $c_{1,0}$
vs $(b-y)_0$ diagram (Fig.\,\ref{fig_c1.RVmembers}). However, the
$c_{1}$ error for this star is large ($\epsilon_{c_{\rm 1}}\sim$ 0.36)
and our classification might be erroneous.

Star \#1823 is one of the stars we classify as a possible RHB/AGB star
because of its location in the $c_{1,0}$ vs $(b-y)_{0}$ diagram along
the RHB/RGB relation from to Anthony-Twarog \& Twarog (1994) (see
Fig.\,\ref{fig.c1.by.zoom.marked}) .

The last discrepant star, \#2320, is a very bright star with $y_0 \sim
16.65$.  In the CMD (Fig.\,\ref{fig.cmd.RVmembers}) it falls far
outside the RGB toward the blue where most of the foreground
contamination is expected.  In the $c_{1,0}$ vs $(b-y)_0$ diagram
(Fig.\,\ref{fig_c1.RVmembers}) it falls on top of the
dwarf star sequence and is therefore classified as a dwarf star. This star
is also found in the proper motion study by Stetson (1980) which will
be discussed in the next section. We  note that Stetson
classified this star as a non-member.

Out of the 27 stars listed as possible radial velocity members in
Kleyna et al. (their Table 2), 10 stars lie within our field and 7 of
those fulfill $y<21$, $\epsilon_{(b-y)}<0.18$, and $\epsilon_{c_{\rm
1}}<0.18$.  Only one (\# 973) is not classified as a RGB member by
us. Its position in the CMD (Fig.\,\ref{fig.cmd.RVmembers}) and in
the $c_{1,0}$ vs $(b-y)_0$ diagram (Fig.\,\ref{fig_c1.RVmembers}A)
shows that it is possibly a RHB/AGB star (see Schuster et al. 2004;
their Fig. 6 for location of RHB/AGB stars).

Finally, we have also cross-correlated the non-Draco members listed in
Armandroff et al. (1995) with our data and find that 13 stars lie
within our field. 11 of those have $y<21$, $\epsilon_{(b-y)}<0.18$, and
$\epsilon_{c_{\rm 1}}<0.18$. We find all of them to be foreground dwarfs
stars.

That the agreement is so good between the radial velocity studies and
our RGB members is not totally unexpected since Kleyna et al. (2002)
especially targeted stars on the red giant branch and our $c_{\rm 1}$
selection criteria select the same type of stars.  However, the result
from this comparison shows that our selection method is very effective
with only two cases where our classification excludes stars that are
possibly members of the Draco dSph galaxy from the radial velocity studies.

\subsubsection{Proper motions}
\label{sect:pm}

\begin{figure}
\resizebox{\hsize}{!}{\includegraphics{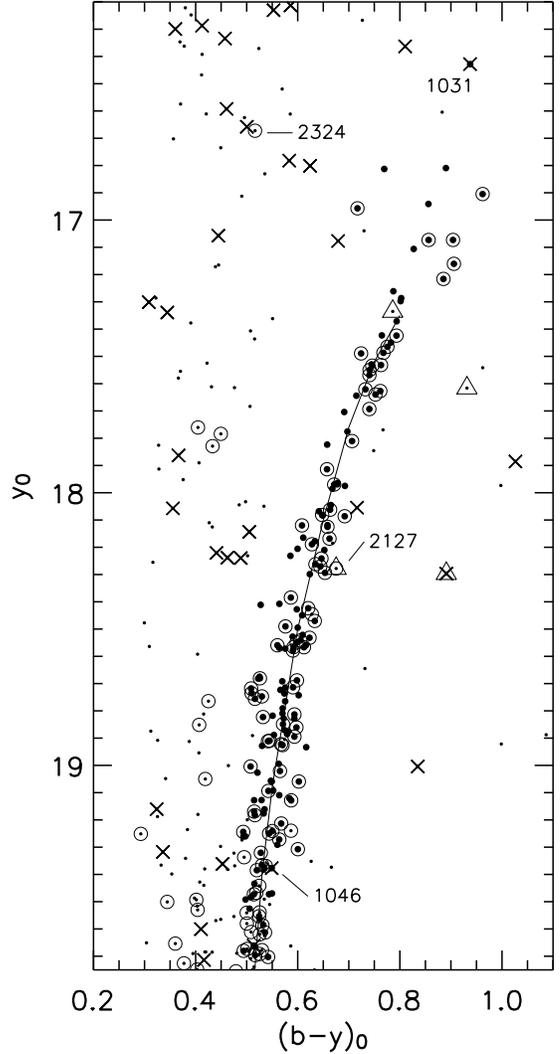}}
\caption{$y_0$ vs $(b-y)_0$ colour-magnitude diagram for the Draco dSph galaxy and
field stars cross-correlated with the proper motion study by Stetson
(1980). All stars have $y<21$, $\epsilon_{(b-y)}<0.18$, and
$\epsilon_{c_{\rm 1}}<0.18$. $\bullet$ mark our RGB stars, $\circ$
indicate stars deemed to be proper motion members based on Stetson
(1980), and $\times$ indicates stars deemed to be non-members in the same
study. Also shown are known carbon stars ($\triangle$). A number of
stars further discussed in the text are marked by their INT numbers.}
\label{fig_c1.PMmembers}
\end{figure}

Stetson (1980) derived proper motions for a large sample of stars in
the Draco dSph galaxy. The proper motions are based on measurements done on
photographic plates. Based on the measured proper motion a membership
probability was derived for each star. Members were then defined as
stars with a probability $P > 0.75$. For details of the derivation of
the proper motions and probabilities the reader is referred to Stetson
(1980).

In Fig. \,\ref{fig_c1.PMmembers} we make a comparison of the
selection of members of the Draco dSph galaxy based on the $c_{\rm 1}$ index and on the
proper motion based probabilities. As can be seen the overall
agreement is good. At the fainter end, below $y_0 \sim$ 19.5, we find
several stars that are proper motion members but not included in our
member list based on the $c_{\rm 1}$ index. This is the direct result
of the cuts applied in the $c_{1,0}$ vs $(b-y)_0$ plane as discussed in
Section 4.2.1. At the brighter end the agreement is nearly perfect. A
number of stars need further commenting.

There is a group of three stars at $(b-y)_0 \sim 0.4$ and $y_0 \sim
17.8$ which we find to be dwarfs while they are included as members in
Stetson (1980). These three stars are, however, found to be
non-members in Armandroff et al. (1995) in agreement with our
classification.

Stars \#1031 and \#1046 are both identified as members using the
$c_{\rm 1}$ index while considered non-members based on their proper
motions. Their membership probabilities, P = 0.71 and 0.53, are,
however, just below the membership cut-off, P = 0.75, used by Stetson
(1980). Star \#1031 is also considered as member of the Draco dSph
galaxy in the radial velocity study by Armandroff et al. (1995). We
therefore feel confident in including these two stars as members of
the Draco dSph galaxy.

Stars \#2127 and \#2324 are both considered members based on their
proper motions while excluded in our study. Star \#2127
is, however, one of the known carbon stars listed in Table 6. Star
\#2324 lies clearly away from the RGB sequence and was identified as a
peculiar UV-bright star in Zinn et al. (1972). Furthermore, Armandroff
et al. (1995) found this star to be a non-member based on radial
velocities.

\subsubsection{Conclusions regarding Draco membership}
Determining Draco membership through selection in the $c_{1}$ vs
$(b-y)$ plane (see Sect. \ref{sect:c1}) has proven to be very
accurate. We find the agreement with other studies using different
methods to be very good and only a few cases are identified where the
different methods disagree. Although other methods for membership
determination can be efficiently used when considering the central
regions of a dwarf galaxy, the $c_{1}$ index method can be used in the
sparsely populated outer regions, where one would expect only a few
members among a dominating foreground contamination.
Table\,\ref{FINALTAB} shows our final list of 169 members of the Draco
dSph galaxy found using the $c_{\rm 1}$ index.

The $c_{1}$ index method used here also has the advantage over other
methods that it provides a way of classifying stars according to their
evolutionary stage without knowing the distance to them. We refer to
the excellent Figure 6 in Schuster et al. (2004) and our
Figs. \,\ref{fig.c1.by.zoom.marked}, \,\ref{fig.c1.by.fullview}, and
\,\ref{fig.cmd.dwarfs.giants} for an illustration of this.

On the other hand, we do not include stars with chemical peculiarities
in our member lists since their colours may vary significantly. This
is for example the case with the carbon stars listed in
Table\,\ref{tab.carbon}.

It is interesting to compare the number of ``discarded'' foreground
stars with expectations from models of the Milky Way stellar
distributions. The Besancon model of stellar population synthesis of
the Milky Way (see Robin et al. 2003) predicts $\sim$ 200 foreground
stars within a 0.067 square degree field (similar to ours) in the
direction of the Draco dSph galaxy with $(b-y)_0 < 1.0$ and $16 < y_0
< 20$.  In our CMD we find a total of 213 stars not identified as RGB
stars and within the magnitude and colour limits given above. This
number is an upper limit since it includes known carbon stars and
possibly some unidentified RGB stars on the faint end of the RGB (see
Fig.  \,\ref{fig.cmd.dwarfs.giants} and discussion in
Sect. \,\ref{sect:c1}). This number is in excellent agreement with the
model prediction. This is further evidence that the method works.

\begin{figure}[t!]
\resizebox{\hsize}{!}{\includegraphics{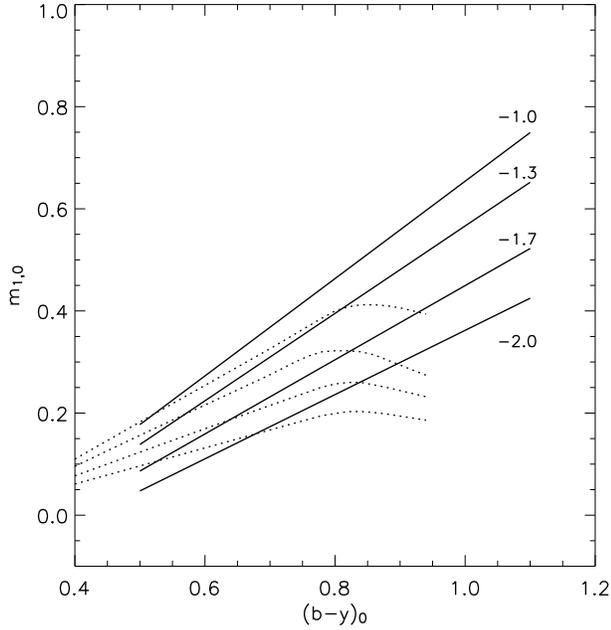}}
\caption{$m_{1,0}$ vs $(b-y)_0$ diagram. Solid lines show isometallicity
calibrations from Hilker (2000) for [Fe/H] = $-1.0$, $-1.3$, $-1.7$, and $-2.0$
dex. Dotted lines show corresponding isometallicity lines from
Anthony-Twarog \& Twarog (1994).}
\label{fig_models_hilk_att}
\end{figure}

\section{Metallicity in the Draco dSph galaxy}
\label{sec.met.in.draco}
To measure the metal content of the stars we use the
$m_{1}=(v-b)-(b-y)$ index. The $m_{\rm1}$ index measures the blocking
by metal lines in the $v$-band and compensates for the slope in the
spectrum measured in a region where metal lines are less prominent
(i.e. in the $b$ and $y$ bands). See e.g. Golay (1974) for more
details on the $m_{\rm1}$ index.

It is known that a molecular CN-band is located within the Str\"omgren
$v$-band (e.g. Gustafsson \& Bell 1979). Stars with high CN-abundances can
therefore mimic stars with higher metallicities. This possibility
should always be taken into account when a scatter towards high metallicities
is seen in metallicity distributions based on the $m_{\rm1}$
index. The metallicity calibrations discussed below are only valid for
CN-weak/normal stars.

\subsection{Metallicity calibrations}
\label{sec.Metcalib}
There are essentially two $m_{1}$ vs $(b-y)$ metallicity calibrations
for RGB stars available in the literature. In the following section we
will review and compare these calibrations in order to establish which
one is the most appropriate for our data.

\begin{figure}[t!]
\resizebox{\hsize}{!}{\includegraphics{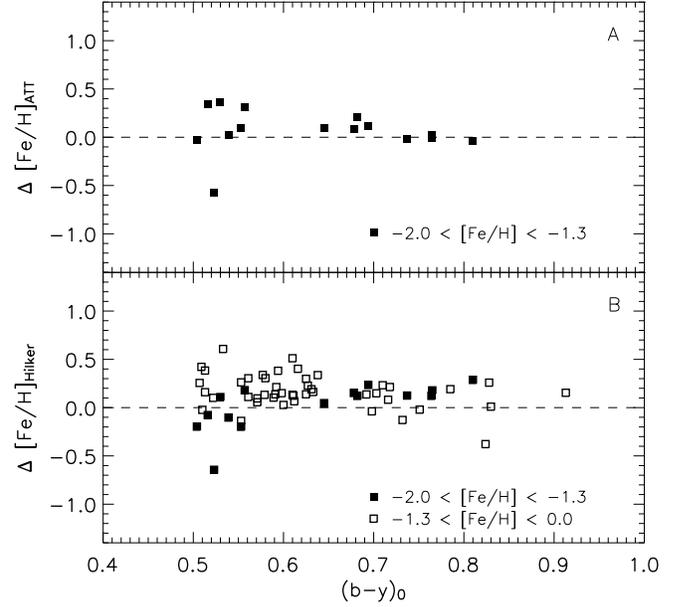}}
\caption{Field star giants from Clem et al. (2004) (their Table 4)
selected by $log\rm{g} < 3.5$ and E$(B-V) < 0.03$. \textbf{A)} shows
the difference between the tabulated metallicity and the metallicity
derived using the Antony-Twarog and Twarog (1994) calibration as a
function of colour. \textbf{B)} shows the equivalent plot using the
Hilker (2000) calibration. }
\label{fig_clemstars_hilk_ATT_by}
\end{figure}

\paragraph{Antony-Twarog \& Twarog (1994)}
present a calibration from $m_{1}$ vs $(b-y)$ to metallicity based on
a sample of  metal-deficient giants. The calibration is therefore
only valid in the metal-poor range $-3.2 < \rm{[Fe/H]} < -1.3$ dex and
within $0.4 < (b-y) < 0.8$.

\paragraph{Hilker (2000)}
presents a metallicity calibration in the the $m_{1}$ vs $(b-y)$ plane
for red giants valid for $-2.0<\rm{[Fe/H]}<0.0$ dex and
$0.5<(b-y)<1.1$. The calibration was derived using red giants from
three globular clusters ($\omega$ Centauri, M55, and M22) as well as a
sample of field giants from Anthony-Twarog \& Twarog (1998). \\

\begin{figure}[t!]
\resizebox{\hsize}{!}{\includegraphics{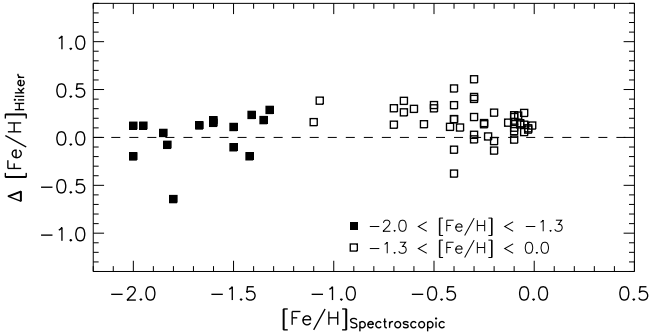}}
\caption{Field giants from Clem et al. (2004) (their Table 4)
selected by $log\rm{g} < 3.5$ and E$(B-V) < 0.03$. Same as
Fig. \,\ref{fig_clemstars_hilk_ATT_by}B but plotted as a function of
metallicities based on spectroscopic investigations.}
\label{fig_clemstars_hilk_fe}
\end{figure}

In Fig. \,\ref{fig_models_hilk_att} we compare isometallicity lines
from Hilker (2000) for [Fe/H] = --1.0, --1.3, --1.7, and --2.0 dex
with those of Antony-Twarog and Twarog (1994) (their Table 4) . In the
region below $(b-y)_{0}$ = 0.8 the two calibrations agree reasonably
well. Although the slopes are slightly different, one could expect to
derive similar metallicities using either of the two calibrations in
this region. A clear difference is, however, the non-linear shape of
the Antony-Twarog and Twarog (1994) models above $(b-y)$ = 0.8 as
compared to the Hilker (2000) calibration. The stellar sample used to
derive the Antony-Twarog and Twarog (1994) calibration contained very
few stars redder than $(b-y)_{0} \sim$ 0.8 and it could be an
indication that the calibration is less reliable above this value.

To investigate this difference further we adopted the field star
sample presented in Clem et al. (2004; their Table 4).  This sample
contains photometry in the four Str\"omgren filters as well as
spectroscopic metallicities for more than 400 stars. Giants were
extracted from the sample by selecting stars with $\log g < 3.5$,
which is the same value adopted by Clem et al. (2004), and E$(B-V) <
0.03$.

By computing $\Delta [\rm{Fe/H}]$, defined as the difference between
the tabulated spectroscopic metallicities and metallicities derived
from the photometric calibrations, we can evaluate the two
calibrations. As a first step we consider all giants within the
overlapping regions of the two calibrations (i.e. $-2.0 < [\rm{Fe/H}]
< -1.3$ and $(b-y) > 0.5$). As expected the two calibrations produce
similar results for stars below $(b-y) = 0.8$
(Fig.\,\ref{fig_clemstars_hilk_ATT_by}). Unfortunately, there is only
one star in this stellar sample with $(b-y) > 0.8$, i.e. where we
expect the difference between the two calibrations to be significant.

Since the Hilker (2000) calibration is valid for higher metallicities,
we have also included stars with $-1.0 < [\rm{Fe/H}] < 0.0$. A number
of these stars have $ (b-y) > 0.8$ and
Fig.\,\ref{fig_clemstars_hilk_ATT_by} shows that there are no visible
trends with colour. The linear approximation of the isometallicity
lines in the Hilker (2000) calibration thus appear valid at the
red end. The metal-poor end will be commented on further in
Sec.\,\ref{sec.complit} when the metallicities of the RGB stars in
Draco dSph galaxy are discussed. We will then argue that a linear
approximation is also valid at the metal-poor end.

While no trend with colour is visible in the Hilker (2000) calibration,
there is a tendency for underestimating ($\Delta[\rm{Fe/H}]_{Hilker} >
0$) the metallicity of stars at the metal-rich end.  This is more
clearly illustrated in Fig.\,\ref{fig_clemstars_hilk_fe}, which shows
the same stellar sample as in Fig.\,\ref{fig_clemstars_hilk_ATT_by}B
but plotted against spectroscopic metallicity rather than colour. A
weak trend can be seen where the metallicity of the most metal-rich
stars are underestimated by, in the mean, $\sim$ 0.2 dex.

\paragraph{Stellar isochrones in the Str\"omgren system}

Recently, Clem et al. (2004) derived new empirical colour-temperature
relations for the Str\"omgren system. These were obtained by
correcting stellar isochrones to fit a sample of field stars including
both red giants and dwarf stars gathered from the literature (the same
field star sample we adopted for the comparison above).

\begin{figure}[t!]
\resizebox{\hsize}{!}{\includegraphics{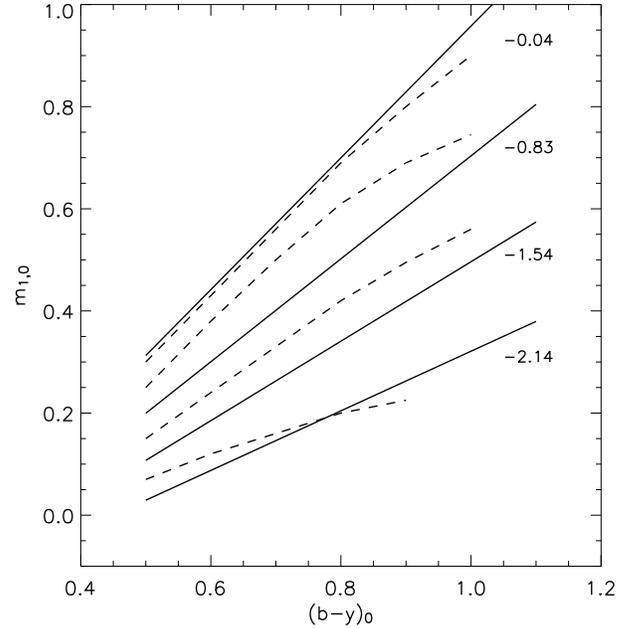}}
\caption{ m$_{1,0}$ vs $(b-y)_0$  diagram. Solid lines show isometallicity
calibrations from Hilker (2000) for [Fe/H] = $-0.04$, $-0.83$, $-1.54$, and $-2.14$
dex. Dashed lines show corresponding isochrones from Clem et
al. (2004; their Fig. 26). }
\label{fig_models_hilk_clem}
\end{figure}

As noted by Clem et al. (2004), a direct comparison in the $m_{1}$ vs
$(b-y)$ plane between their calibrated isochrones and the metallicity
calibration by Hilker (2000) shows significant differences.  In
Fig. \,\ref{fig_models_hilk_clem} we illustrate this by showing Clem
et al. (2004) isochrones for [Fe/H] = --0.83, --1.54, and --2.14 dex
(from their Fig. 26) overlaid on the equivalent isometallicity lines
from Hilker (2000). The models agree well at the metal-poor and
metal-rich end (i.e. [Fe/H] = --2.0 and 0.0 dex). There is, however, a
strong discrepancy at intermediate metallicities with a difference
between the two models as large as $\sim$0.5 dex at [Fe/H] $\sim$
--0.8 dex. Note that the Clem et al. (2004) isochrones fall above the
equivalent Hilker (2000) isometallicity lines. This would result in an
even stronger underestimate of the metallicities of the field star
sample compared to the Hilker (2000) calibration. This result is
surprising since the field star sample we consider is the same sample
of stars used by Clem et al. (2004) to correct their models.

\paragraph{Summary}
We have shown that the Hilker (2000) and Antony-Twarog and Twarog
(1994) calibrations give similar results in their overlapping
metallicity region and below $(b-y) = 0.8$. There are no trends with
colour visible in the Hilker (2000) calibration and that the linear
shape of the isometallicity lines is a good approximation, at least
for the metal-rich end.  The Hilker (2000) calibration also has the
advantage of extending up to solar metallicity and it is capable of
deriving correct metallicities for an independent field star
sample. We therefore adopt the Hilker (2000) calibration to derive
metallicities for the RGB stars in the Draco dSph galaxy.

We note, however, that there are unexplained and significant differences
between the Hilker (2000) and Clem et al. (2004) models, which should be
understood before adopting these calibrations at the metal rich end. A
discussion of this issue is ongoing with Clem et al. (Private
communication).

\subsection{Metallicity determination}

\begin{figure}[t]
\resizebox{\hsize}{!}{\includegraphics{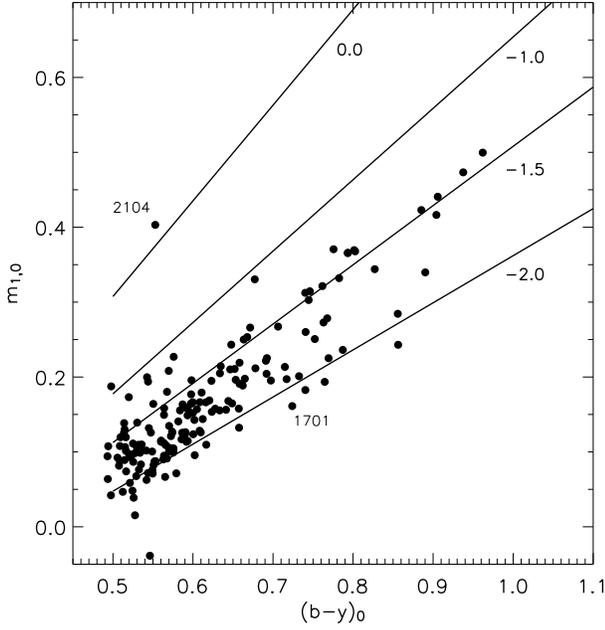}}
\caption{$m_{1,0}$ vs $(b-y)_0$ diagram showing all our RGB stars in the  Draco
dSph galaxy. Overlaid are isometallicity lines from Hilker (2000) for [Fe/H]
= 0.0, --1.0, --1.5, and --2.0 dex.}
\label{fig_clemstars_hilkcalib.ps}
\end{figure}

Figure \,\ref{fig_clemstars_hilkcalib.ps} shows that the RGB members in
the  Draco dSph galaxy
form a metal-poor population between --2.0 and --1.5 dex. 
For these stars we derive [Fe/H] using the calibrations from Hilker (2000). 
These are listed in Table\,\ref{FINALTAB}.

Again we note that star \#2104 deviates considerably from the rest of
the RGB sample (compare Fig.\ref{fig.cmd.dwarfs.giants}C). If indeed
an RGB member, the metallicity for this star is close to solar, which
is considerably higher than what we see for the rest of the stars in
the Draco dSph galaxy. The most likely interpretation is that this
star is a foreground dwarf star misidentified as RGB star (see
Sect. \,\ref{sect:c1}).

\subsection{Metallicity errors}
\label{sect:feherrors}
\begin{figure}[t]
\resizebox{\hsize}{!}{\includegraphics{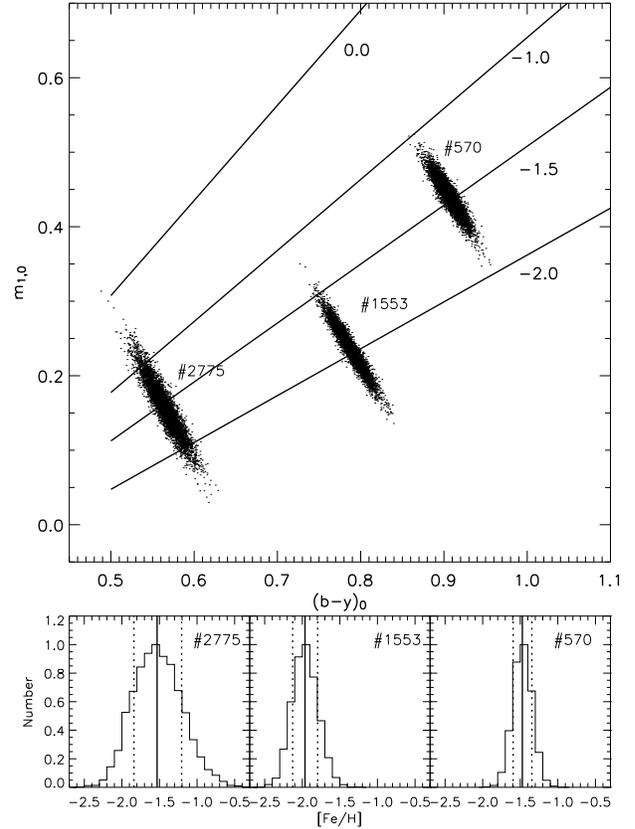}}
\caption{Illustration of how the errors in the metallicities are derived
(see Sect.\,\ref{sect:feherrors}).
Upper panel shows the $m_{1,0}$ vs $(b-y)_0$ diagram with
Monte Carlo simulations for three stars (\#570, \#1553, and
\#2775). Overlaid are isometallicity lines from Hilker (2000) for
[Fe/H]=0.0, --1.0, --1.5, and --2.0 dex. Lower panels show
corresponding metallicity distribution functions for the synthetic
stars. Solid lines indicate the metallicity of the real stars. Dotted
lines indicate the upper- and lower sixtile in the distribution.  }
\label{fig_MCErrors.ps}
\end{figure}

To handle the error propagation from photometric errors for individual
magnitudes, as shown in Fig. \,\ref{fig.errors}, to errors in [Fe/H]
we use a Monte Carlo simulation of the data. This was done in the
following way: for each magnitude of a given star (i.e. $u$, $v$, $b$,
and $y$) we generate 5000 new synthetic magnitudes. These were
randomly drawn from a Gaussian probability distribution with a
standard deviation equal to the error ($\epsilon$) in that magnitude.
New $m_{1}$ indices and $(b-y)$ colours are then calculated for each
synthetic star and we derive the corresponding metallicities.

Figure \,\ref{fig_MCErrors.ps} shows three examples of
how the synthetic stars generated for star \#570, \#1553, and \#2775
are distributed in the $m_{1,0}$ vs $(b-y)_0$ plane.  It is interesting to
note how the errors are coupled so that the synthetic stars spread
out perpendicularly to the isometallicity lines. This spread is mostly
driven by the error in the $b$ filter.

We also present the corresponding metallicity distribution functions
for the three stars. The upper and lower sixtile of the distributions
(which is equivalent to 1 $\sigma$ in the case of a Gaussian
distribution) are chosen to represent our errors.

It is worth noting that the metallicity distributions for the
synthetic stars are not fully symmetric. This is not surprising since
the spacing of the isometallicity lines in the $m_{1,0}$ vs
$(b-y)_0$ plane is not constant. The errors defined above are therefore
not necessarily symmetric around the original [Fe/H] value. However,
since the difference of the upper and lower sixtile is typically less
than a few percent, we give the final [Fe/H] errors as half the
distance between the upper and lower sixtile.

\begin{figure}[t]
\resizebox{\hsize}{!}{\includegraphics{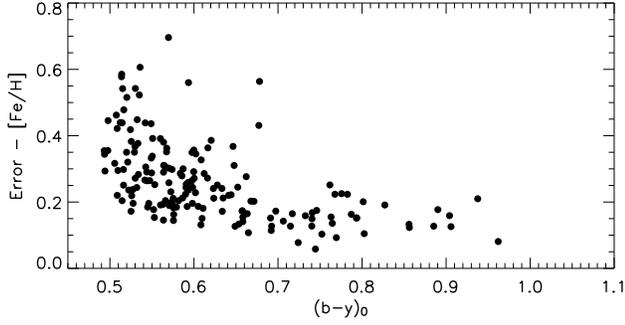}}
\caption{Final errors in [Fe/H] as a function of $(b-y)_0$ for all our 
RGB stars in the Draco dSph galaxy.}
\label{fig_FeErrors.ps}
\end{figure}

Figure \,\ref{fig_FeErrors.ps} shows the final errors in [Fe/H],
derived as explained above, as a function of $(b-y)_0$.

\subsection{Comparison with spectroscopically derived metallicities}
\label{sec.complit}
Because of the large, and to a large extent unexplained, discrepancies
between the different metallicity calibrations discussed above, it is
useful to compare our derived metallicities with independent
measurements, e.g. spectroscopically determined metallicities. Such a
comparison could also reveal any systematic shifts or trends in our
data.

Abundance studies of stars in the nearby dSph galaxies are, however,
difficult and time consuming due to the faintness of even the
brightest of their RGB stars. In Table\,\ref{tab_FE_lit} we have
collected metallicities for RGB stars in the Draco dSph galaxy available in
the literature.

\begin{figure}[t!]
\resizebox{\hsize}{!}{\includegraphics{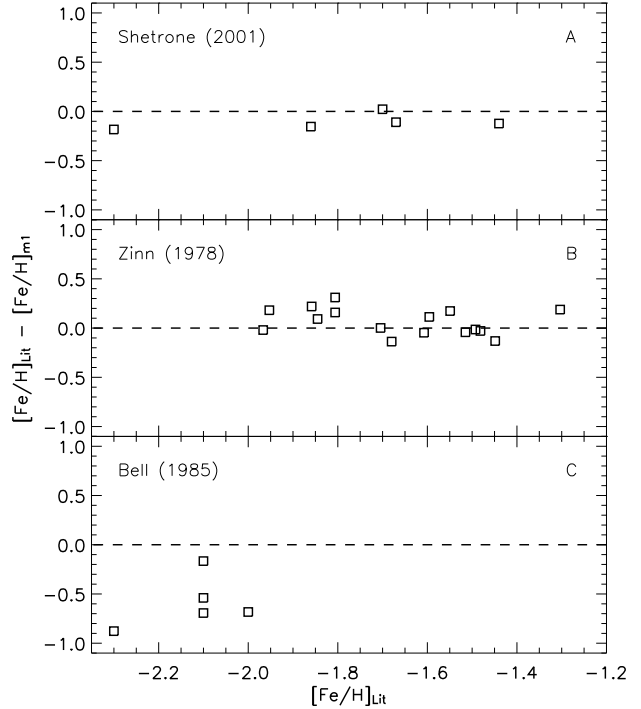}}
\caption{Comparison between our derived metallicities ([Fe/H]$_{m1}$)
and values from the literature([Fe/H]$_{Lit}$): \textbf{A)} Shetrone et
al. (2001a). \textbf{B)} Zinn et al. (1978) corrected to the scale by Carreta
\& Gratton (1997) as described in Sect. \,\ref{sec.complit}. \textbf{C)} Bell
 (1985).}
\label{Literature_Fe}
\end{figure}

The most recent data set, Shetrone et al. (2001a), is based on
high-resolution spectra and applies a detailed abundance analysis
based on equivalent width measurements.  Figure\,\ref{Literature_Fe}A
shows the difference between the metallicities derived by us using the
$m_{1}$ index and the values measured by Shetrone et al. (2001a) for
five stars which are common to the studies. The agreement with the
Shetrone et al. (2001a) values are remarkably good with a standard
deviation $\sigma = 0.08$ and a mean difference of 0.1 dex.  In Table
\,\ref{tab_FE_lit} a sixth star is included (\#1701) for which
Shetrone et al. (2001a) quotes [Fe/H] = --2.97 dex, or $\sim$ 0.84 dex
lower than our value. The Shetrone et al. (2001a) value is well below
the limit of --2.0 dex for which the Hilker (2000) metallicity
calibration is valid and therefore we did not include this star in
Fig.\,\ref{Literature_Fe}. In our data star \#1701 falls below the
--2.0 dex isometallicity line but not by much \textbf{(see
Fig.\,\ref{fig_clemstars_hilkcalib.ps})}. This star is also included in
the study by Zinn (1978) (see below) and the metallicity found by him
is closer to the value we find than to that by Shetrone et
al. (2001a).

We note that Fulbright et al.\,(2004) have studied this star in
detail. They derive a metallicity similar to that found by Shetrone et
al.\,(2001a) but with an effective temperature ($T_{\rm eff}$) that is
$\sim 100$K larger than used in Shetrone et al.\,(2001a). We have
calculated $T_{\rm eff}$ for all six stars in common with Shetrone et
al.\,(2001a) using the calibrations in Alonso et al.\,(1999). For five
of the stars (excluding \#1701) the agreement between $T_{\rm eff}$
derived from $(b-y)$ and $(u-b)$ is good.  For those three stars that
also have $(B-V)$ from Stetson (photometry available at {\tt
  http://cadcwww.hia.nrc.ca/standards/}) the derived $T_{\rm eff}$
also agree very well. However, all our derivations differ from the
values in Shetrone et al.\,(2001a) such that we derive $T_{\rm eff}$
that are $\sim$100 -- 200 K lower. For \#1701 the agreement between
$T_{\rm eff}$ derived from $(b-y)$ and $(u-b)$ is less good than for
the other stars. Also, we find a temperature that is 150 K higher for
$(u-b)$, in agreement with Fulbright et al.\,(2004). However, a change
of $T_{\rm eff}$ of 150 K does not make a large enough change in
[Fe/H] to explain the differences between the different
studies. Neither does a change in $\log g$ (see Fulbright et al.,
2004). Hence, there appears to be no easy explanation for the large
discrepancy between our photometric metallicity and the metallicity
derived from high resolution spectroscopy. A cautionary remark on the
usage of plane parallel atmospheres in the derivation of elemental
abundances for stars with $\log g<2.0$ (which is the case for the star
considered here) has recently been added by Heiter \& Eriksson
(2006). However, to ascertain if this is the cause of the observed
discrepancy needs to be further investigated by spectral modelling and
is beyond the scope of this paper.

The most extensive dataset is provided by Zinn (1978). His estimates
of the metallicities are based on the $Q(3880)$ index (Zinn, 1978 and
Zinn \& Searl, 1976). Before comparing our derived metallicities to
those of Zinn (1978) we apply the corrections to the Zinn \& West
scale described in Carretta \& Gratton
(1997). Figure\,\ref{Literature_Fe}B shows the comparison between our
data and the corrected metallicities derived by Zinn (1978). Again, we
see that our derived values are in good agreement with those of Zinn
(1978). No trends can be seen and the standard deviation is $\sigma =
0.13$.

Several of the stars included in the Shetrone et al. (2001a) and/or the
Zinn (1978) studies have $(b-y)_{0} > 0.8$ (see Table
\,\ref{tab_FE_lit}). Plotting the metallicity differences as a
function of $(b-y)_{0} > 0.8$ rather than metallicity shows that there
are no trends with colour. This is encouraging since it confirms the
validity of the linear shape of the Hilker (2000) metallicity lines at
the metal-poor and red end, a region where the two metallicity
calibrations discussed in Sec. \,\ref{sec.Metcalib} differed
significantly.

Finally, Fig.\,\ref{Literature_Fe}C shows a comparison with abundances
measured by Bell (1985). In this case we see a rather large average
shift, on the order of $\sim$ 0.5 dex. Five of the stars in both Bell
(1985) and our data sets are also found in either Shetrone (2001a)
(three stars) and/or Zinn (1978) data (five stars). In both cases the
Bell (1985) values are consistently lower by $\sim$ 0.5 dex compared
to the other studies.

In conclusion, the agreement between our derived metallicities with
those derived using spectroscopic methods is very good. Our errors in
[Fe/H] are consistent with the spreads seen in
Fig.\,\ref{Literature_Fe}A and B which show that we are not underestimating
the uncertainties in our derived metallicities.  The absence of large
shifts between our photometric [Fe/H] and those from other studies
show that our data are free from large systematic errors.

\subsection{Draco metallicity distribution}
\begin{figure*}[t!]
\resizebox{\hsize}{!}{\includegraphics{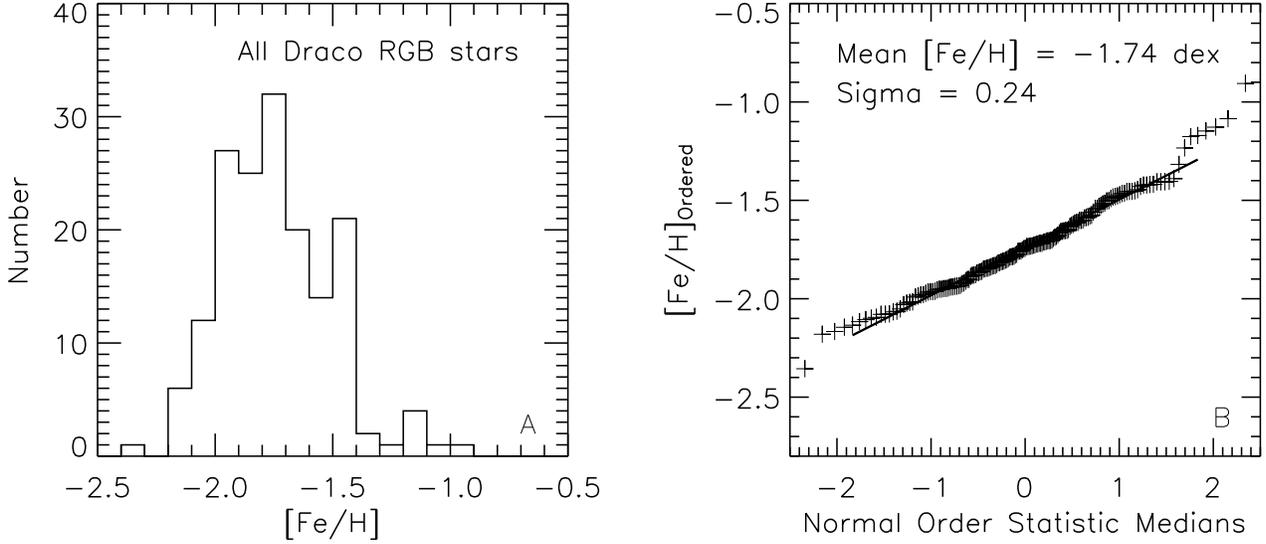}}
\caption{\textbf{A)} metallicity distribution functions for all the RGB
stars in the Draco dSph galaxy
(169 stars and bin size = 0.10 dex).  \textbf{B)} shows
corresponding probability plot assuming a Gaussian distribution. Solid
line shows a linear fit to the data with a slope of 0.24 and the
intersection with the y-axis at --1.74.}
\label{Draco_fe_histogram1}
\end{figure*}

Figure \,\ref{Draco_fe_histogram1}A presents the full metallicity
distribution in the Draco dSph galaxy which is one of the main results
of this paper. The  metallicity ranges from below [Fe/H] = --2
dex to [Fe/H] $\sim$ --1.5 dex with a small tail towards higher
metallicities.

Figure \,\ref{Draco_fe_histogram1}B presents the corresponding normal
probability function (Devore 2000) which shows that the metallicity
distribution is indeed consistent with being a single Gaussian
distribution (i.e. the data points fall on a straight line). The data
points only deviate significantly from a straight line at the upper
right hand corner of the plot.  This corresponds to the tail of
metal-rich stars seen in the histogram.  The solid line represents a
linear least square fit to the data (excluding the outer five values
at both ends) and give the mean [Fe/H] = --1.74 dex (intersection of
the fitted line with x-axis = 0) and a $\sigma$ of the distribution
equal to 0.24 dex (slope of the fitted line).

\begin{figure*}[t!]
\resizebox{\hsize}{!}{\includegraphics{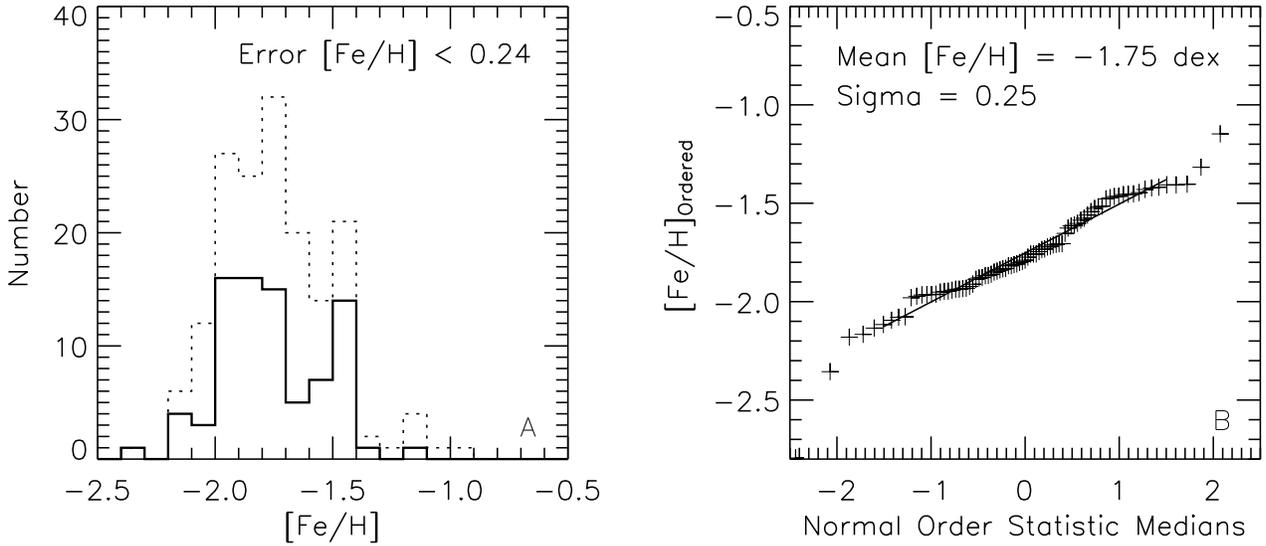}}
\caption{\textbf{A)} Metallicity distribution functions for all members
of the RGB in the Draco dSph galaxy
with error in $\rm{[Fe/H]} < 0.24$ dex (85 stars and bin size
= 0.10). Dotted line shows distribution for all members of the Draco dSph galaxy (same as
in Fig. \,\ref{Draco_fe_histogram1}).  \textbf{B)} Corresponding
normal probability plot assuming a Gaussian distribution. Solid line
shows a linear fit to the data with a slope of 0.25 and the
intersection with the y-axis at --1.75.}
\label{Draco_fe_histogram2}
\end{figure*}

Since stars with large photometric errors (especially at the blue end)
will cause large uncertainties in the derived metallicities as
discussed above, we extracted a subsample of stars with errors in
$\rm{[Fe/H]}$ less than 0.24 dex. Roughly half of our stars fulfill
this criterion (85 out of 169).  The resulting metallicity
distribution and corresponding probability function for this subsample
are shown in Figure\,\ref{Draco_fe_histogram2}.  The derived mean
[Fe/H] and $\sigma$ for this distribution (--1.75 dex and 0.25 dex,
respectively) are almost identical to those of the full sample.

There is a hint of a double peak in the distribution in
Fig.\,\ref{Draco_fe_histogram1} with a main peak at [Fe/H] $\sim$
--1.8 dex and a second, smaller peak at [Fe/H] $\sim$ --1.4 dex. This
secondary peak can also be seen in the full metallicity distribution
in Fig. \,\ref{Draco_fe_histogram1}A . A KMM test (e.g. Ashman \& Bird
1994) gives a probability of $\sim$6\% that the observed distribution
is drawn from a single Gaussian distribution. The corresponding
probability for the subsample with low errors in their metallicities
is $\sim$20\%.

The corresponding probability function shows a slight ``s'' shape
which is consistent with a bimodal distribution but this indication is
weak.

We also note that the tail extending towards higher metallicities
disappears when only stars with small errors are considered.  This may
suggest that the tail is a result of our photometric errors rather
than a true intrinsic feature in the Draco dSph galaxy. A closer look
at the stars constituting this metal-rich tail shows that while most
of them have errors larger than the average error, they are not
extreme. Many of the stars fall just outside the error cut-off for our
subsample (see above and Table\,\ref{FINALTAB}).

It is also interesting to note that a subsample of stars which have
errors in [Fe/H] that are larger than $ 0.4$ have distribution with an
identical mean [Fe/H] = --1.74 dex and $\sigma$ = 0.24 dex as found
for the full sample and also for the subsample with small errors. This
indicates that it is not the errors in the derived metallicities that drive the
width of the distribution but a true intrinsic metallicity spread in
the Draco dSph galaxy.

The derived mean metallicity of --1.74 dex with a spread of 0.24 dex
that we find for the full sample is comparable with recent studies
using different methods. Lehnert et al. (1992), Dolphin (2002),
Shetrone et al. (2001a), and Bellazzini et al. (2002) all find similar mean
metallicities and evidence for a significant intrinsic metallicity
spread comparable to ours.

\begin{figure}[t!]
\resizebox{\hsize}{!}{\includegraphics{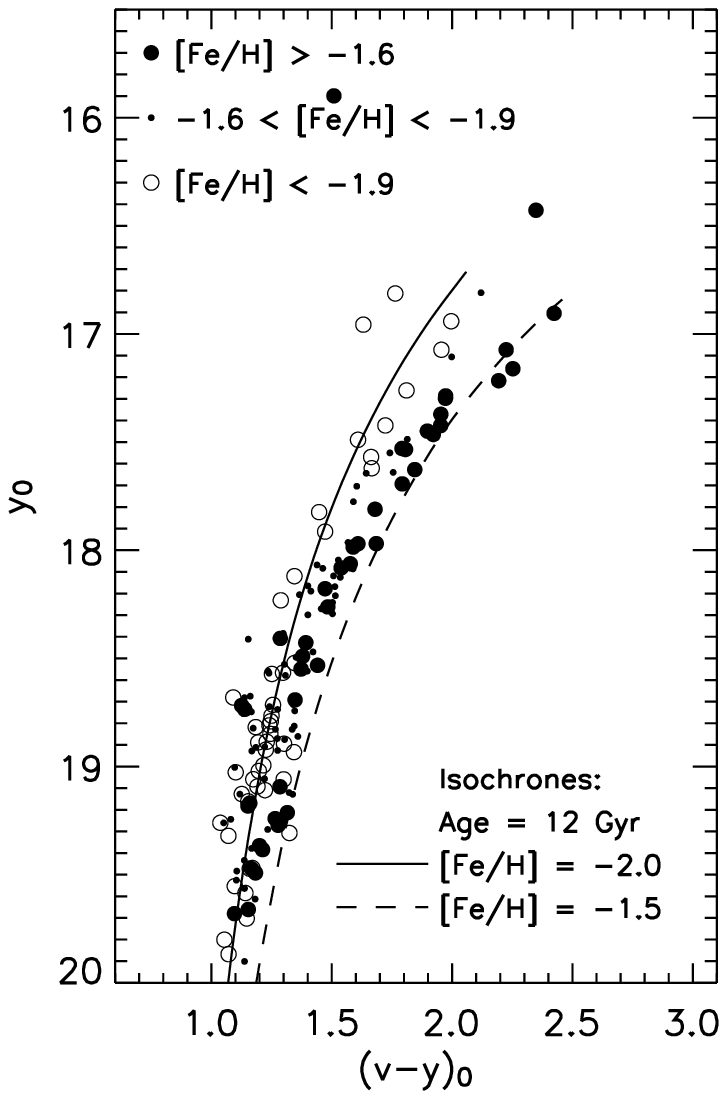}}
\caption{$y_0$ vs $(b-y)_0$ colour-magnitude diagram for all our RGB members for
the Draco dSph galaxy. Large
$\bullet$ indicate metal-rich stars with $\rm{[Fe/H]} > -1.6$
dex. Small $\bullet$ indicates intermediate metallicity stars with
$-1.9 < \rm{[Fe/H]} < -1.6$ dex. $\circ$ indicates metal-poor stars
with $\rm{[Fe/H]} < -1.9$ dex. 12 Gyr isochrones by Bergbusch \&
VandenBerg (2001) with $uvby$ colour transformations as described by Clem
et al (2004) are overlaid. The solid line indicate an isochrone with [Fe/H] = --2 dex and the
dashed line with [Fe/H] = --1.5 dex.}
\label{fig_fe_cmd.ps}
\end{figure}

Shown in Fig.\,\ref{fig_fe_cmd.ps} is a CMD of our sample of RGB stars
in the Draco dSph galaxy subdivided, according to their metallicities,
into three bins with $[\rm{Fe/H}] > -1.6$ dex, $-1.9 < \rm{[Fe/H]} <
-1.6$ dex, and $\rm{[Fe/H]} < -1.9$ dex. The metallicity bins contain
45, 77, and 47 stars, respectively. On the lower part of the RGB the
metal-rich and metal-poor stars are intermingled probably as a
consequence of the larger photometric errors at fainter
magnitudes. Above $y \sim 18.5$, however, the metal-poor and
metal-rich stars are separated in colour. The width of the RGB is well
explained with a pure metallicity spread assuming an old stellar
population as found by e.g. Grillmair et al (1998), Grebel (2001), and
Dolphin (2002). This is illustrated in the figure by the overlaid
isochrones (by Bergbusch \& VandenBerg (2001) with $uvby$ colour
transformations as described by Clem et al (2004)) with age = 12 Gyr
and [Fe/H] = --2 and --1.5 dex. The isochrones have been shifted
assuming a distance modulus of $(m-M)_{0} = 19.40$ (Bonanos et
al. 2004).

A similar CMD for the Carina dSph galaxy is shown in Koch et
al. (2006), their Fig. 10. In that case there is no clear relation
between metallicity and colour on the RGB as a result of a significant
age spread. This is the well known age-metallicity degeneracy, and
illustrates the importance of age-independent metallicities when
deriving accurate metallicity distribution functions.

\begin{table*}[t!]
\caption[]{Abundances from the literature for stars in the Draco
dSph galaxy. Columns 1 and 2 gives ID number from Baade \& Swope (1961) and
from this study, respectively; Col. 3-6 give our Str\"omgren
photometry; Col. 7 the metallicity derived in this paper using the
$m_{1}$ index; Col. 8-10 metallicities, effective temperatures, and
$\log g$ from Shetrone et al. (2001a); Col 11 metallicities from Bell
(1985) and Col. 12 metallicities from Zinn (1978).}
\label{tab_FE_lit}
\begin{tabular}{llllllllllll}
\hline\hline 
\noalign{\smallskip}
BS  & \multicolumn{6}{l}{INT}    & \multicolumn{3}{l}{S012}                 &{B85} & Z78\\
id  & id & $y_{0}$ & $(b-y)_{0}$ & $m_{1,0}$ & $c_{1,0}$ & [Fe/H]$_{\rm m_{\rm 1}}$ & [Fe/H] & T$_{\rm eff}$ & $\log (g)$  &[M/H] & [Fe/H]  \\
\noalign{\smallskip}
\hline
\noalign{\smallskip}
G   &  1297 & 17.569 &  0.740 &  0.183 &  0.518 & --2.08 &       &      &       &--2.5  & --2.02  \\
11  &  1940 & 17.532 &  0.763 &  0.273 &  0.516 & --1.72 & --1.7 & 4475 &  0.80 &       & --1.77  \\
24  &  2082 & 17.073 &  0.856 &  0.243 &  0.585 & --2.12 & --2.3 & 4290 &  0.80 &--2.7  & --1.98  \\
45  &  2097 & 17.621 &  0.733 &  0.201 &  0.535 & --1.96 &       &      &       &       & --1.98  \\
49  &  1954 & 16.957 &  0.717 &  0.197 &  0.550 & --1.93 &       &      &       &--2.1  &         \\
72  &  2106 & 18.294 &  0.653 &  0.196 &  0.511 & --1.71 &       &      &       &       & --1.81  \\
119 &  1701 & 17.489 &  0.724 &  0.161 &  0.427 & --2.13 &--2.97 & 4370 &  0.15 &       & --2.09  \\
249 &  1988 & 17.216 &  0.885 &  0.423 &  0.457 & --1.48 &       &      &       &       & --1.72  \\
267 &  2366 & 17.073 &  0.904 &  0.416 &  0.578 & --1.56 &--1.67 & 4180 &  0.60 &--2.1  & --1.82  \\
286 &  2334 & 17.693 &  0.740 &  0.312 &  0.481 & --1.45 &       &      &       &       & --1.71  \\
297 &  2421 & 17.811 &  0.706 &  0.267 &  0.561 & --1.54 &       &      &       &       & --1.88  \\
343 &  1772 & 17.550 &  0.741 &  0.260 &  0.520 & --1.71 &--1.86 & 4475 &  0.90 &       & --1.90  \\
361 &  1112 & 17.424 &  0.793 &  0.365 &  0.498 & --1.41 &       &      &       &--2.1  &         \\
473 &  2501 & 17.465 &  0.776 &  0.371 &  0.489 & --1.32 &--1.44 & 4400 &  0.90 &--2.0  & --1.68  \\
506 &  2194 & 17.914 &  0.657 &  0.158 &  0.495 & --1.94 &       &      &       &       & --2.01  \\
536 &  1142 & 16.905 &  0.962 &  0.500 &  0.491 & --1.42 &       &      &       &--2.3  &         \\
562 &  1553 & 17.161 &  0.906 &  0.441 &  0.538 & --1.47 &       &      &       &--2.2  & --1.74  \\
576 &  1073 & 16.941 &  0.856 &  0.284 &  0.549 & --1.95 &       &      &       &       & --2.10  \\
581 &  1110 & 17.627 &  0.762 &  0.321 &  0.525 & --1.49 &       &      &       &       & --1.54  \\ 
\noalign{\smallskip}
\hline
\end{tabular}
\end{table*}

\subsection{Spatial metallicity distribution}
It is a known fact that many dSph galaxies in the Local Group show
population gradients (see e.g. Harbeck et al. 2001 and references
therein; Koch et al. 2006). In the case of the Draco dSph galaxy,
population gradients have been found in at least two photometric
studies. Bellazini et al. (2002) find that the red HB stars are more
centrally concentrated than the blue HB stars in their sample and
interpret this as central concentration of more metal rich and/or
younger stars. A similar trend is found by Winnick (2003). Using
spectroscopic metallicities derived from the CaII triplet on a sample
of 95 members of the Draco dSph galaxy, Winnick (2003) finds that the
most metal-rich stars are centrally concentrated.

A conflicting result is presented by Cioni \& Habing (2005), who find that
if the age of the stellar population in the Draco dSph galaxy is indeed old, the
metallicity increases outwards.

Since we have derived age-independent metallicities for individual
stars in a clean Draco sample (i.e. we are not biased by the
age-metallicity degeneracy effect), we should be able to detect
spatial metallicity gradients in our data if they are present. 

\begin{figure}[t!]
\resizebox{\hsize}{!}{\includegraphics{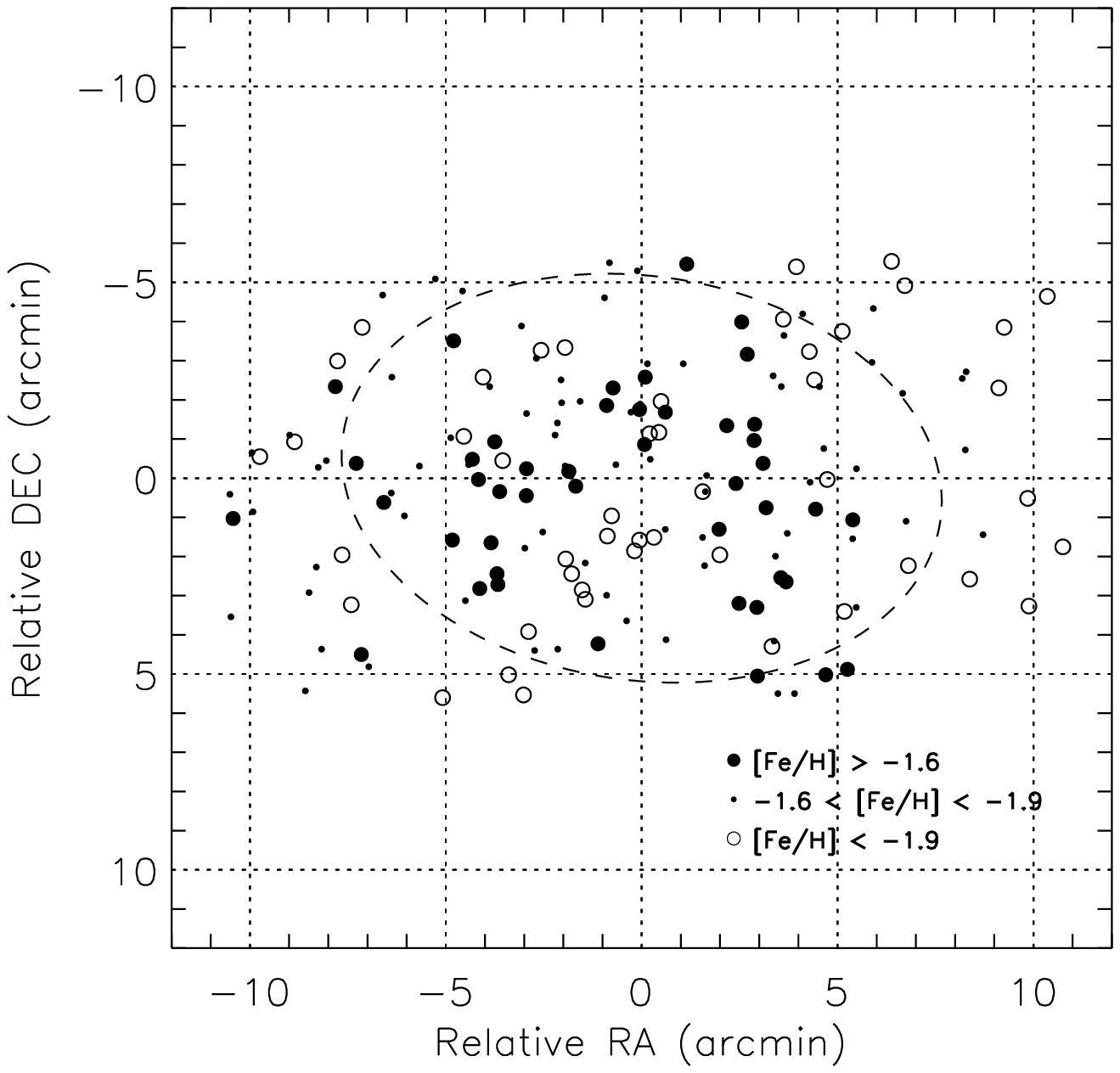}}
\caption{Spatial distribution of RGB stars in the Draco dSph galaxy. Large $\bullet$
indicate metal-rich stars with $\rm{[Fe/H]} > -1.6$ dex, small
$\bullet$ stars with $-1.6 < \rm{[Fe/H]} < -1.9$ dex, and $\circ$
metal-poor stars with $\rm{[Fe/H]} < -1.9$ dex. The dashed ellipse
indicates the core radius from Irwin \& Hatzidimitriou (1995),
r$_{core} = 7.7$ arcmin.  }
\label{fig_XY.ps}
\end{figure}

Figure\,\ref{fig_XY.ps} shows the spatial distribution of the RGB
stars in the Draco dSph galaxy subdivided into the same three
metallicity bins. The figure shows a similar trend to what is shown in
Fig. 2.20 in Winnick (2003), with the metal-rich stars being more
centrally concentrated than the metal poor-stars.

\begin{figure}[t!]
\resizebox{\hsize}{!}{\includegraphics{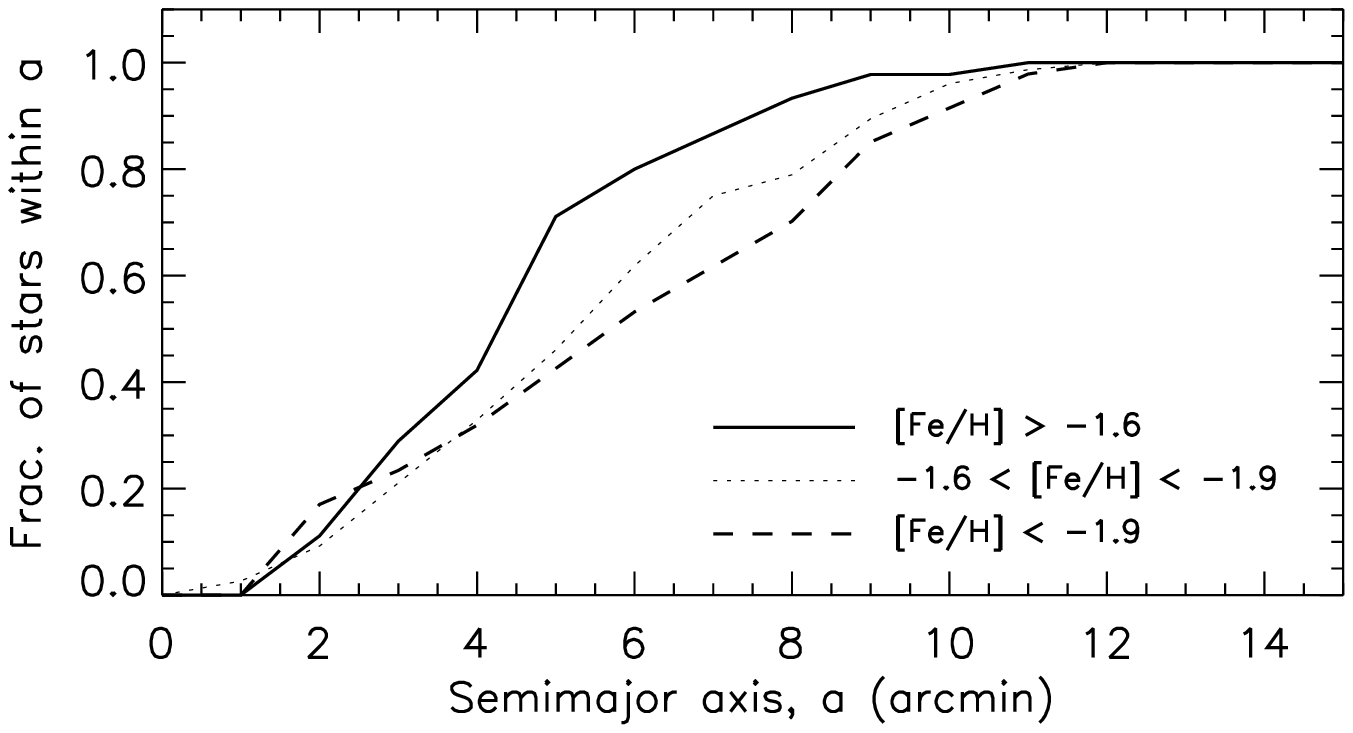}}
\caption{Normal cumulative distribution for semimajor axis, $a$. Solid
line indicates metal-rich stars with $\rm{[Fe/H]} > -1.6$ dex, dotted
line stars with $-1.6 < \rm{[Fe/H]} < -1.9$ dex, and dashed line
metal-poor stars with $\rm{[Fe/H]} < -1.9$ dex.}
\label{fig_XY_ell.ps}
\end{figure}

In Fig.\,\ref{fig_XY_ell.ps} we show the corresponding cumulative
distributions for the different metallicity bins. The fraction of
stars within an ellipse with semi-major axis, \textit{a}, and ellipticity =
0.33 (Irwin \& Hatzidimitriou 1995) is plotted against semi-major axis,
$a$. The metal-rich stars show a more centrally concentrated
distribution than the metal-poor stars. A two-sided Kolmogorov-Smirnov
test gives a probability of less than 2\% that the metal-poor and
metal-rich spatial distributions are the same. For consistency, we also
include the cumulative distribution for the stars with intermediate
metallicities. As expected this curve falls between the metal-rich and
metal-poor distributions.

\section{Conclusions and summary}
\label{sect:summary}

The aim of this study is to provide a robust method for the
identification of RGB stars that are members of a dSph galaxy and
determination of metallicities for individual stars along the RGB of a
dSph galaxy. For the first task it is necessary to have a method that
can distinguish between a foreground dwarf star and an RGB star in the
dSph. The second task requires a metallicity sensitive index. The
Str\"omgren photometric system ( $uvby$) provides both. More
specifically we have:

\begin{itemize}

 \item proven the usefulness of the Str\"omgren $c_{\rm 1}$ index
 in discriminating between RGB stars in dSph galaxies and foreground
 dwarf stars.
\\
 \item presented a clean RGB sample for the Draco dSph galaxy.
\\
 \item investigated the available metallicity calibrations for the
 Str\"omgren $m_{\rm 1}$ index 
\\
 \item derived the metallicity distribution function based on
age-independent metallicities for individual stars in the inner part
of the Draco dSph galaxy  \\
\item shown that the more metal-rich RGB stars in the Draco dSph galaxy
  are more centrally
  concentrated than the metal-poor RGB stars.\\
\end{itemize}

We also include cross correlation with the following available data
sets: Stetson (1980) broad band photometry in B and V, radial velocity
studies by Armandroff et al. (1995) and Kleyna et al. (2002), and
Baade \& Swope (1961) list of variable stars in the Draco dSph galaxy.

Membership has been determined using the Str\"omgren $c_{\rm 1}$-colour
index.  Comparison with alternative methods of membership
determination (e.g. proper motion and radial velocity measurements) 
show that our membership classification
agrees very well with other methods. We are therefore
confident that the same method for membership determination,
i.e. selection in the $c_{\rm 1}$ vs $(b-y)$-plane, can now be applied
to the sparsely populated outer regions of dSphs. 

In addition to the ability to identify dSph members using the $c_{\rm
1}$ index, the Str\"omgren system provides the possibility to derive
individual and age-independent metallicities for RGB stars using
the $m_{\rm 1}$ index.  Since our metallicity determination is age
independent, our results are not limited by the age-metallicity
degeneracy (which is the case for most other photometric metallicity
studies).  

A review of the existing Str\"omgren metallicity calibrations for
giant stars has led us to use the calibration by Hilker (2000).
Applying it to our clean sample of members of the Draco dSph galaxy we
present individual metallicities for 169 stars, the largest
independent sample so far for the Draco dSph galaxy. The
photometrically derived metallicities agree very well with
high-resolution spectroscopic determinations (i.e. Shetrone et al.,
2001a) and with earlier results from spectral indices (Zinn, 1978).

The metallicity distribution function we obtain is consistent with a
single Gaussian distribution with a mean metallicity [Fe/H] = --1.74
dex and a $\sigma$ = 0.25 dex and with a small tail of more metal-rich
stars. 

Although the data presented in this paper only include the central
regions of the Draco dSph galaxy, we have investigated the spatial metallicity
distribution and find evidence for a central concentration of more 
metal-rich stars.

\begin{acknowledgement}

We thank L Lindegren at Lund Observatory for valuable advice regarding
the error propagation and the data analysis.  DF thanks J Lewis at the
Wide Field Survey unit at IoA, Cambridge, for invaluable help during
the data reduction.  SF and GG thank the Swedish Royal Society for a
collaborative grant that made it possible for SF and DF to visit
Cambridge.  SF acknowledges a visiting scientist grant from ESO/Chile
to visit DF and during which visit important parts of the work were
finalized.  SF is a Royal Swedish Academy of Sciences Research Fellow
supported by a grant from the Knut and Alice Wallenberg Foundation.
These observations have been funded by the Optical Infrared
Coordination network (OPTICON), a major international collaboration
supported by the Research Infrastructures Program of the European
Commissions Fifth Framework Program.
\end{acknowledgement}

\begin{sidewaystable*}
\begin{minipage}[t][180mm]{\textwidth}
\caption{RGB stars in the Draco dSph galaxy. Column 1 gives the ID number from this
study; Col. 2 gives the designation from Stetson's photometric catalog
(available at {\tt http://cadcwww.hia.nrc.ca/standards/}); Cols. 3 and
4 give the pixel position on the CCD; Cols. 5 and 6 give the position
on the sky, Cols. 7 -- 10 give our Str\"omgren photometry and Col. 11
the metallicity derived in this paper using the $m_{1}$ index. Column
12 gives the membership probability based on proper motions given in
Stetson (1980). Finally Col. 13 gives radial velocities from Kleyna et
al. (2002).}
\label{FINALTAB}
\centering
\begin{tabular}{llrrlllllllllll}
\hline\hline             

INT ID & Other ID &  x & y & R.A. (2000) & Dec (2000) & $y\pm \epsilon_y$  &  $b\pm \epsilon_{b}$  & $v\pm \epsilon_v$ &  $u\pm \epsilon_u$ & [Fe/H]$_{\rm m_{\rm 1}}$ & P & $V_{\rm r}$\\

\hline
  53 & &   676.192 &    69.349 &   4.5446477 &   1.0113240 &    19.801 $\pm$   0.022 &    20.347 $\pm$   0.014 &    20.854 $\pm$   0.049 &    21.706 $\pm$   0.171  & -2.79$\pm$0.20 &       &         \\
  92 &S-366 &  1838.474 &   125.208 &   4.5444217 &   1.0094626 &    17.423 $\pm$   0.018 &    18.188 $\pm$   0.023 &    19.146 $\pm$   0.012 &    20.559 $\pm$   0.079  & -2.09$\pm$0.14    &       & -304.75        \\
 162 & &   909.711 &   228.023 &   4.5441618 &   1.0109553 &    19.260 $\pm$   0.041 &    19.758 $\pm$   0.029 &    20.297 $\pm$   0.032 &    21.225 $\pm$   0.106  & -2.03$\pm$0.37 &       &         \\
\hline																 
\end{tabular}															 
\vfill																 
\end{minipage}															 
\end{sidewaystable*}

\Online

\begin{table}
      \caption[]{List of standard stars used (from Schuster \& Nissen
1988). Column 1 gives the star ID; Col. 2--5 give the magnitudes and colours.}
         \label{tab.standards}
         \begin{tabular}{lllll}
            \hline
            \noalign{\smallskip}
            ID & $y_{0}$ & $(b - y)_{0}$ & $m_{1,0}$  & $c_{1,0}$  \\
            \noalign{\smallskip}
            \hline
            \noalign{\smallskip}
HD 33449     & 8.488  &   0.423  &   0.201  &   0.273 \\  
HD 46341     & 8.616  &   0.366  &   0.145  &   0.248 \\  
HD 51754     & 9.000  &   0.375  &   0.144  &   0.290 \\  
HD 64090     & 8.279  &   0.428  &   0.110  &   0.126 \\   
HD 75530     & 9.167  &   0.443  &   0.254  &   0.257 \\ 
HD 81408     & 9.610  &   0.560  &   0.478  &   0.210 \\ 
HD 88371     & 8.414  &   0.407  &   0.186  &   0.329 \\ 
HD 107853    & 9.081  &   0.317  &   0.157  &   0.483 \\  
HD 107583    & 9.33   &   0.375  &   0.183  &   0.319 \\ 
HD 108754    & 9.006  &   0.435  &   0.217  &   0.254 \\ 
HD 118659    & 8.827  &   0.422  &   0.196  &   0.244 \\    
HD 123265    & 8.348  &   0.504  &   0.356  &   0.348 \\    
HD 131653    & 9.506  &   0.442  &   0.226  &   0.253 \\     
HD 132475    & 8.555  &   0.401  &   0.063  &   0.285 \\  
HD 134088    & 7.992  &   0.392  &   0.137  &   0.255 \\ 
HD 134439    & 9.058  &   0.484  &   0.224  &   0.165 \\  
HD 134440    & 9.419  &   0.524  &   0.297  &   0.173 \\ 
HD 137303    & 8.774  &   0.611  &   0.610  &   0.178 \\ 
HD 138648    & 8.137  &   0.504  &   0.358  &   0.290 \\ 
HD 149414    & 9.611  &   0.476  &   0.202  &   0.162 \\  
HD 161770    & 9.696  &   0.489  &   0.036  &   0.301 \\ 
HD 163810    & 9.635  &   0.423  &   0.114  &   0.199 \\ 
HD 175179    & 9.072  &   0.384  &   0.146  &   0.268 \\ 
HD 175617    & 10.130 &   0.441  &   0.208  &   0.281 \\ 
HD 220769    & 9.31   &   0.340  &   0.140  &   0.310 \\ 
G 9 -031     & 10.823 &   0.398  &   0.158  &   0.224 \\ 
G 9 -036     & 11.934 &   0.381  &   0.124  &   0.195 \\     
G 14 -024    & 12.822 &   0.509  &   0.123  &   0.094 \\ 
G 14 -039    & 12.828 &   0.587  &   0.267  &   0.153 \\     
G 14 -045    & 10.803 &   0.587  &   0.517  &   0.115 \\ 
G 63 -026    & 12.183 &   0.328  &   0.085  &   0.277 \\    
DM -14 4454  & 10.332 &   0.565  &   0.469  &   0.192 \\ 
DM -14 3322  & 10.394 &   0.377  &   0.131  &   0.220 \\ 
DM -13 2948  & 9.439  &   0.426  &   0.256  &   0.221 \\ 
DM -13 3834  & 10.685 &   0.415  &   0.098  &   0.183 \\  
DM -12 2669  & 10.230 &   0.229  &   0.094  &   0.490 \\ 
DM -9 3102   & 10.479 &   0.425  &   0.202  &   0.206 \\ 
DM -8 4501   & 10.591 &   0.452  &   0.032  &   0.274 \\ 
DM -5 3063   & 9.734  &   0.568  &   0.461  &   0.182 \\ 	    
DM -5 2678   & 10.654 &   0.296  &   0.155  &   0.422 \\ 
DM -5 3763   & 10.239 &   0.579  &   0.546  &   0.241 \\ 
DM -4 3208   & 9.998  &   0.311  &   0.048  &   0.373 \\ 
DM +25 1981  & 9.317  &   0.237  &   0.103  &   0.489 \\ 
             \noalign{\smallskip}
            \hline
         \end{tabular}
   \end{table}

\begin{sidewaystable*}
\begin{minipage}[t][180mm]{\textwidth}
\caption{Draco RGB stars. Column 1 gives the ID number from this
study; Col. 2 gives the designation from Stetson's photometric catalog
(available at {\tt http://cadcwww.hia.nrc.ca/standards/}); Cols. 3 and
4 give the pixel position on the CCD; Cols. 5 and 6 give the position
on the sky, Cols. 7 -- 10 give our Str\"omgren photometry and Col. 11
the metallicity derived in this paper using the $m_{1}$ index. Column
12 gives the membership probability based on proper motions given in
Stetson (1980). Finally Col. 13 gives radial velocities from Kleyna et
al. (2002).}
\label{FINALTAB}
\centering
\begin{tabular}{llrrlllllllllll}
\hline\hline             

INT ID & Other ID &  x & y & R.A. (2000) & Dec (2000) & $y\pm \epsilon_y$  &  $b\pm \epsilon_{b}$  & $v\pm \epsilon_v$ &  $u\pm \epsilon_u$ & [Fe/H]$_{\rm m_{\rm 1}}$ & P & $V_{\rm r}$\\

\hline
  53 & &   676.192 &    69.349 &   4.5446477 &   1.0113240 &    19.801 $\pm$   0.022 &    20.347 $\pm$   0.014 &    20.854 $\pm$   0.049 &    21.706 $\pm$   0.171  & -2.79$\pm$0.20 &       &         \\
  92 & S-366 &  1838.474 &   125.208 &   4.5444217 &   1.0094626 &    17.423 $\pm$   0.018 &    18.188 $\pm$   0.023 &    19.146 $\pm$   0.012 &    20.559 $\pm$   0.079  & -2.09$\pm$0.14 &       &         \\
 162 & &   909.711 &   228.023 &   4.5441618 &   1.0109553 &    19.260 $\pm$   0.041 &    19.758 $\pm$   0.029 &    20.297 $\pm$   0.032 &    21.225 $\pm$   0.106  & -2.03$\pm$0.37 &       &         \\
 167 & &   404.888 &   231.823 &   4.5441723 &   1.0117639 &    19.127 $\pm$   0.044 &    19.656 $\pm$   0.035 &    20.253 $\pm$   0.017 &    21.254 $\pm$   0.124  & -1.99$\pm$0.38 &       &         \\
 243 &S-348 &  1698.858 &   324.059 &   4.5438347 &   1.0096923 &    19.585 $\pm$   0.050 &    20.118 $\pm$   0.043 &    20.726 $\pm$   0.026 &    21.692 $\pm$   0.241  & -1.94$\pm$0.47 &       &         \\
 266 &S-346 &  1420.164 &   360.157 &   4.5437412 &   1.0101408 &    19.027 $\pm$   0.032 &    19.548 $\pm$   0.033 &    20.128 $\pm$   0.014 &    21.131 $\pm$   0.056  & -2.02$\pm$0.33 &       &         \\
 338 & &   741.701 &   444.257 &   4.5435209 &   1.0112317 &    18.450 $\pm$   0.019 &    19.058 $\pm$   0.043 &    19.833 $\pm$   0.036 &    20.979 $\pm$   0.400  & -1.69$\pm$0.35 &       & -291.42 \\
 372 & &   539.093 &   506.334 &   4.5433431 &   1.0115584 &    18.231 $\pm$   0.026 &    18.817 $\pm$   0.033 &    19.520 $\pm$   0.014 &    20.630 $\pm$   0.084  & -1.90$\pm$0.29 &       &         \\
 374 &S-338 &  1135.257 &   519.022 &   4.5432787 &   1.0106027 &    18.828 $\pm$   0.022 &    19.401 $\pm$   0.024 &    20.094 $\pm$   0.032 &    21.167 $\pm$   0.088  & -1.82$\pm$0.24 &       &         \\
 378 & &  1500.894 &   509.574 &   4.5432892 &   1.0100154 &    17.964 $\pm$   0.087 &    18.642 $\pm$   0.072 &    19.532 $\pm$   0.026 &    20.758 $\pm$   0.097  & -1.72$\pm$0.62 &       &         \\
 379 &S-337 &  1465.066 &   521.863 &   4.5432544 &   1.0100733 &    18.165 $\pm$   0.018 &    18.776 $\pm$   0.022 &    19.566 $\pm$   0.009 &    20.718 $\pm$   0.126  & -1.63$\pm$0.18 &       &         \\
 570 & &   606.555 &   793.316 &   4.5424771 &   1.0114594 &    17.261 $\pm$   0.017 &    18.048 $\pm$   0.030 &    19.072 $\pm$   0.011 &    20.629 $\pm$   0.038  & -1.96$\pm$0.18 &       &         \\
 571 &S-311 &   814.854 &   798.362 &   4.5424533 &   1.0111253 &    18.210 $\pm$   0.027 &    18.862 $\pm$   0.032 &    19.725 $\pm$   0.017 &    20.979 $\pm$   0.128  & -1.62$\pm$0.25 &       &         \\
 574 & &  1898.897 &   789.996 &   4.5424275 &   1.0093836 &    18.885 $\pm$   0.016 &    19.465 $\pm$   0.023 &    20.115 $\pm$   0.016 &    21.100 $\pm$   0.083  & -2.17$\pm$0.19 &       &         \\
 576 & &  1402.435 &   803.061 &   4.5424128 &   1.0101818 &    19.057 $\pm$   0.023 &    19.604 $\pm$   0.027 &    20.277 $\pm$   0.029 &    21.281 $\pm$   0.217  & -1.67$\pm$0.28 &       &         \\
 593 &S-302 &  2016.722 &   842.690 &   4.5422630 &   1.0091954 &    18.521 $\pm$   0.011 &    19.131 $\pm$   0.019 &    19.867 $\pm$   0.023 &    20.964 $\pm$   0.128  & -1.94$\pm$0.15 &       &         \\
 639 & &  1799.953 &   932.962 &   4.5420027 &   1.0095463 &    19.564 $\pm$   0.022 &    20.089 $\pm$   0.041 &    20.701 $\pm$   0.038 &    21.668 $\pm$   0.257  & -1.83$\pm$0.40 &       &         \\
 649 &S-301 &  1402.435 &   803.061 &   4.5424128 &   1.0101818 &    19.057 $\pm$   0.023 &    19.604 $\pm$   0.027 &    20.277 $\pm$   0.029 &    21.281 $\pm$   0.217  & -1.67$\pm$0.28 &       &         \\
 593 & &  1547.761 &   945.233 &   4.5419779 &   1.0099521 &    18.528 $\pm$   0.003 &    19.118 $\pm$   0.031 &    19.831 $\pm$   0.010 &    20.948 $\pm$   0.046  & -1.88$\pm$0.23 &       &         \\
 729 &S-298 &  1057.142 &  1023.448 &   4.5417652 &   1.0107428 &    17.776 $\pm$   0.018 &    18.473 $\pm$   0.026 &    19.365 $\pm$   0.016 &    20.715 $\pm$   0.048  & -1.88$\pm$0.18 &       &         \\
 752 & &   733.721 &  1046.272 &   4.5417094 &   1.0112629 &    18.911 $\pm$   0.024 &    19.453 $\pm$   0.029 &    20.096 $\pm$   0.011 &    21.122 $\pm$   0.086  & -1.80$\pm$0.27 &  0.94 &         \\
 760 & &   416.900 &  1032.803 &   4.5417633 &   1.0117707 &    17.975 $\pm$   0.013 &    18.667 $\pm$   0.007 &    19.585 $\pm$   0.048 &    20.942 $\pm$   0.091  & -1.71$\pm$0.13 &       &         \\
 798 & &   129.392 &  1076.583 &   4.5416422 &   1.0122328 &    17.449 $\pm$   0.028 &    18.232 $\pm$   0.036 &    19.346 $\pm$   0.017 &    20.951 $\pm$   0.067  & -1.52$\pm$0.22 &       &         \\
 800 &S-293 &  1693.595 &  1081.475 &   4.5415611 &   1.0097210 &    17.824 $\pm$   0.014 &    18.481 $\pm$   0.026 &    19.271 $\pm$   0.008 &    20.495 $\pm$   0.056  & -2.08$\pm$0.19 &       &         \\
 810 & &   400.710 &  1086.382 &   4.5416021 &   1.0117984 &    18.888 $\pm$   0.014 &    19.442 $\pm$   0.030 &    20.083 $\pm$   0.026 &    21.063 $\pm$   0.111  & -1.96$\pm$0.27 &       &         \\
 864 & &  1156.473 &  1170.984 &   4.5413156 &   1.0105871 &    18.910 $\pm$   0.009 &    19.454 $\pm$   0.020 &    20.131 $\pm$   0.020 &    21.135 $\pm$   0.090  & -1.61$\pm$0.19 &  0.93 &         \\
 877 & &   107.980 &  1179.928 &   4.5413308 &   1.0122702 &    17.297 $\pm$   0.025 &    18.099 $\pm$   0.034 &    19.270 $\pm$   0.008 &    20.942 $\pm$   0.142  & -1.42$\pm$0.22 &       &         \\
 884 &S-286 &  1443.377 &  1192.707 &   4.5412374 &   1.0101264 &    19.260 $\pm$   0.023 &    19.754 $\pm$   0.032 &    20.312 $\pm$   0.033 &    21.233 $\pm$   0.116  & -1.84$\pm$0.37 &       &         \\
 902 &S-285 &  1469.530 &  1214.630 &   4.5411701 &   1.0100847 &    19.469 $\pm$   0.028 &    20.019 $\pm$   0.032 &    20.640 $\pm$   0.018 &    21.577 $\pm$   0.287  & -2.05$\pm$0.30 &       &         \\
 907 &392 &   877.310 &  1217.484 &   4.5411873 &   1.0110371 &    18.532 $\pm$   0.017 &    19.155 $\pm$   0.033 &    19.973 $\pm$   0.011 &    21.121 $\pm$   0.179  & -1.59$\pm$0.26 &  0.85 &         \\
 910 &386 &  1002.836 &  1242.023 &   4.5411077 &   1.0108360 &    19.121 $\pm$   0.035 &    19.704 $\pm$   0.025 &    20.444 $\pm$   0.051 &    21.506 $\pm$   0.141  & -1.65$\pm$0.31 &       &         \\
 922 &S-283 &  1605.947 &  1236.966 &   4.5410967 &   1.0098660 &    18.791 $\pm$   0.035 &    19.361 $\pm$   0.009 &    20.037 $\pm$   0.019 &    21.083 $\pm$   0.109  & -1.91$\pm$0.19 &       &         \\
 939 & &  1777.872 &  1262.273 &   4.5410123 &   1.0095899 &    18.736 $\pm$   0.037 &    19.310 $\pm$   0.031 &    20.011 $\pm$   0.025 &    21.133 $\pm$   0.085  & -1.78$\pm$0.31 &       &         \\
 956 & &  1998.828 &  1290.207 &   4.5409174 &   1.0092350 &    19.092 $\pm$   0.010 &    19.644 $\pm$   0.019 &    20.284 $\pm$   0.028 &    21.360 $\pm$   0.164  & -1.95$\pm$0.18 &       &         \\
 980 & &    22.130 &  1324.380 &   4.5408974 &   1.0124120 &    16.809 $\pm$   0.014 &    17.699 $\pm$   0.028 &    18.929 $\pm$   0.072 &    20.750 $\pm$   0.147  & -1.82$\pm$0.19 &       &         \\
 993 & &   768.016 &  1350.871 &   4.5407887 &   1.0112163 &    18.542 $\pm$   0.024 &    19.147 $\pm$   0.020 &    19.909 $\pm$   0.022 &    21.063 $\pm$   0.094  & -1.74$\pm$0.20 &  0.95 &         \\
 996 &S-274 &  1680.806 &  1349.157 &   4.5407548 &   1.0097482 &    18.119 $\pm$   0.005 &    18.777 $\pm$   0.023 &    19.626 $\pm$   0.008 &    20.936 $\pm$   0.094  & -1.76$\pm$0.15 &       &         \\
1006 &S-273 &  1756.981 &  1351.336 &   4.5407443 &   1.0096257 &    18.572 $\pm$   0.022 &    19.147 $\pm$   0.019 &    19.823 $\pm$   0.016 &    20.861 $\pm$   0.155  & -1.97$\pm$0.18 &       &         \\
1024 &544 &  1441.549 &  1367.591 &   4.5407100 &   1.0101336 &    18.926 $\pm$   0.041 &    19.496 $\pm$   0.020 &    20.201 $\pm$   0.023 &    21.314 $\pm$   0.228  & -1.72$\pm$0.28 &  0.88 &         \\
1031 &414 &   542.374 &  1355.997 &   4.5407825 &   1.0115788 &    16.428 $\pm$   0.044 &    17.366 $\pm$   0.038 &    18.777 $\pm$   0.009 &    20.673 $\pm$   0.100  & -1.45$\pm$0.21 &  0.71 & -300.47 \\
\hline																 
\end{tabular}															 
\vfill																 
\end{minipage}															 
\end{sidewaystable*}

\addtocounter{table}{-1}

\begin{sidewaystable*}
\begin{minipage}[t][180mm]{\textwidth}
\caption{Continued.}
\label{FINALTAB}
\centering
\begin{tabular}{llrrlllllllllll}
\hline\hline             

INT ID & Other ID &  x & y & R.A. (2000) & Dec (2000) & $y\pm \epsilon_y$  &  $b\pm \epsilon_{b}$  & $v\pm \epsilon_v$ &  $u\pm \epsilon_u$ & [Fe/H]$_{\rm m_{\rm 1}}$ & P & $V_{\rm r}$\\

\hline
1032 & &   561.460 &  1379.797 &   4.5407100 &   1.0115488 &    18.692 $\pm$   0.118 &    19.261 $\pm$   0.020 &    20.039 $\pm$   0.085 &    21.222 $\pm$   0.307  & -1.23$\pm$0.74 &       &         \\
1041 & &    24.222 &  1403.295 &   4.5406590 &   1.0124109 &    19.176 $\pm$   0.042 &    19.709 $\pm$   0.034 &    20.346 $\pm$   0.043 &    21.376 $\pm$   0.065  & -1.75$\pm$0.41 &       &         \\
1046 &542 &  1497.098 &  1403.805 &   4.5405979 &   1.0100452 &    19.377 $\pm$   0.031 &    19.926 $\pm$   0.049 &    20.575 $\pm$   0.044 &    21.569 $\pm$   0.125  & -1.85$\pm$0.45 &  0.53 &         \\
1051 & &   658.005 &  1404.104 &   4.5406322 &   1.0113945 &    19.004 $\pm$   0.044 &    19.512 $\pm$   0.045 &    20.101 $\pm$   0.014 &    21.083 $\pm$   0.202  & -1.78$\pm$0.50 &  0.97 &         \\
1072 & &   268.593 &  1415.126 &   4.5406141 &   1.0120198 &    18.495 $\pm$   0.031 &    19.095 $\pm$   0.042 &    19.847 $\pm$   0.035 &    21.003 $\pm$   0.122  & -1.74$\pm$0.38 &       &         \\
1073 &576 &   242.520 &  1427.791 &   4.5405765 &   1.0120618 &    16.941 $\pm$   0.017 &    17.797 $\pm$   0.023 &    18.937 $\pm$   0.030 &    20.626 $\pm$   0.086  & -1.95$\pm$0.13 &       & -282.44 \\
1079 &166 &   887.666 &  1445.687 &   4.5404973 &   1.0110265 &    19.169 $\pm$   0.036 &    19.684 $\pm$   0.042 &    20.329 $\pm$   0.025 &    21.347 $\pm$   0.134  & -1.46$\pm$0.47 &  0.97 &         \\
1084 &377 &  1088.096 &  1454.157 &   4.5404639 &   1.0107045 &    18.718 $\pm$   0.025 &    19.226 $\pm$   0.019 &    19.843 $\pm$   0.011 &    20.829 $\pm$   0.113  & -1.59$\pm$0.23 &  0.94 &         \\
1101 &372 &  1195.812 &  1495.384 &   4.5403342 &   1.0105323 &    19.661 $\pm$   0.074 &    20.175 $\pm$   0.041 &    20.814 $\pm$   0.032 &    21.793 $\pm$   0.231  & -1.48$\pm$0.64 &  0.31 &         \\
1104 & &   424.930 &  1493.678 &   4.5403705 &   1.0117711 &    18.736 $\pm$   0.038 &    19.245 $\pm$   0.021 &    19.874 $\pm$   0.018 &    20.885 $\pm$   0.081  & -1.50$\pm$0.31 &  0.97 &         \\
1110 &581 &   109.851 &  1492.713 &   4.5403848 &   1.0122763 &    17.627 $\pm$   0.028 &    18.389 $\pm$   0.040 &    19.472 $\pm$   0.023 &    21.080 $\pm$   0.111  & -1.49$\pm$0.26 &  0.83 & -301.95 \\
1112 &361 &  1271.348 &  1491.454 &   4.5403433 &   1.0104107 &    17.424 $\pm$   0.008 &    18.217 $\pm$   0.027 &    19.376 $\pm$   0.019 &    21.033 $\pm$   0.093  & -1.41$\pm$0.15 &  0.95 & -287.25 \\
1142 &536 &  1596.831 &  1519.234 &   4.5402455 &   1.0098875 &    16.905 $\pm$   0.012 &    17.867 $\pm$   0.016 &    19.328 $\pm$   0.015 &    21.280 $\pm$   0.181  & -1.42$\pm$0.08 &  0.94 & -300.67 \\
1159 & &  1744.505 &  1545.400 &   4.5401597 &   1.0096503 &    18.178 $\pm$   0.014 &    18.811 $\pm$   0.023 &    19.650 $\pm$   0.015 &    20.912 $\pm$   0.095  & -1.58$\pm$0.18 &       &         \\
1191 & &   997.708 &  1583.382 &   4.5400767 &   1.0108532 &    19.094 $\pm$   0.031 &    19.636 $\pm$   0.026 &    20.378 $\pm$   0.034 &    21.471 $\pm$   0.193  & -1.13$\pm$0.32 &  0.96 &         \\
1202 &427 &   447.718 &  1574.937 &   4.5401239 &   1.0117366 &    18.082 $\pm$   0.009 &    18.729 $\pm$   0.045 &    19.620 $\pm$   0.015 &    20.924 $\pm$   0.049  & -1.42$\pm$0.32 &  0.97 & -301.76 \\
1223 &153 &  1269.643 &  1620.402 &   4.5399537 &   1.0104165 &    18.490 $\pm$   0.014 &    19.066 $\pm$   0.018 &    19.869 $\pm$   0.009 &    21.075 $\pm$   0.052  & -1.15$\pm$0.17 &  0.97 &         \\
1297 &G &   669.692 &  1663.150 &   4.5398479 &   1.0113825 &    17.569 $\pm$   0.011 &    18.309 $\pm$   0.025 &    19.232 $\pm$   0.016 &    20.673 $\pm$   0.054  & -2.08$\pm$0.15 &  0.97 &         \\
1324 &194 &   791.767 &  1664.778 &   4.5398383 &   1.0111864 &    18.262 $\pm$   0.024 &    18.897 $\pm$   0.027 &    19.746 $\pm$   0.012 &    21.059 $\pm$   0.105  & -1.53$\pm$0.23 &  0.97 &         \\
1341 &146 &  1037.402 &  1716.289 &   4.5396729 &   1.0107925 &    18.567 $\pm$   0.019 &    19.179 $\pm$   0.038 &    19.935 $\pm$   0.019 &    21.056 $\pm$   0.156  & -1.84$\pm$0.30 &  0.96 &         \\
1356 &219 &   623.093 &  1732.477 &   4.5396404 &   1.0114591 &    18.470 $\pm$   0.032 &    19.104 $\pm$   0.030 &    19.892 $\pm$   0.013 &    21.183 $\pm$   0.164  & -1.86$\pm$0.25 &  0.96 &         \\
1359 &171 &   964.446 &  1725.615 &   4.5396476 &   1.0109102 &    18.861 $\pm$   0.044 &    19.458 $\pm$   0.019 &    20.220 $\pm$   0.029 &    21.329 $\pm$   0.166  & -1.65$\pm$0.26 &  0.94 &         \\
1360 &172 &   963.479 &  1737.057 &   4.5396132 &   1.0109119 &    18.895 $\pm$   0.103 &    19.488 $\pm$   0.040 &    20.196 $\pm$   0.054 &    21.309 $\pm$   0.225  & -1.95$\pm$0.61 &  0.94 &         \\
1365 &195 &   751.090 &  1738.848 &   4.5396156 &   1.0112535 &    18.579 $\pm$   0.017 &    19.169 $\pm$   0.031 &    19.886 $\pm$   0.021 &    20.995 $\pm$   0.065  & -1.86$\pm$0.25 &  0.97 &         \\
1444 & &  2012.633 &  1795.190 &   4.5393939 &   1.0092242 &    17.529 $\pm$   0.013 &    18.274 $\pm$   0.006 &    19.321 $\pm$   0.007 &    20.798 $\pm$   0.062  & -1.51$\pm$0.06 &       &         \\
1458 &350 &  1558.326 &  1819.533 &   4.5393395 &   1.0099564 &    18.189 $\pm$   0.012 &    18.817 $\pm$   0.036 &    19.602 $\pm$   0.013 &    20.837 $\pm$   0.137  & -1.83$\pm$0.26 &  0.94 & -289.00 \\
1553 &562 &   824.661 &  1044.222 &   4.5417123 &   1.0111169 &    17.161 $\pm$   0.017 &    18.066 $\pm$   0.024 &    19.413 $\pm$   0.016 &    21.298 $\pm$   0.169  & -1.47$\pm$0.12 &  0.97 & -295.02 \\
1567 &558 &  1010.729 &  1159.957 &   4.5413556 &   1.0108211 &    18.566 $\pm$   0.022 &    19.158 $\pm$   0.021 &    19.864 $\pm$   0.016 &    20.964 $\pm$   0.087  & -1.95$\pm$0.19 &  0.91 &         \\
1647 &122 &  1336.791 &  1904.973 &   4.5390902 &   1.0103151 &    17.969 $\pm$   0.016 &    18.646 $\pm$   0.063 &    19.654 $\pm$   0.017 &    21.016 $\pm$   0.136  & -1.09$\pm$0.45 &       &         \\
1654 &L &   794.353 &  1910.505 &   4.5390944 &   1.0111881 &    18.168 $\pm$   0.038 &    18.831 $\pm$   0.035 &    19.681 $\pm$   0.010 &    21.044 $\pm$   0.052  & -1.79$\pm$0.28 &  0.97 &         \\
1701 &119 &  1385.871 &  1923.999 &   4.5390306 &   1.0102365 &    17.489 $\pm$   0.009 &    18.213 $\pm$   0.010 &    19.098 $\pm$   0.021 &    20.411 $\pm$   0.065  & -2.13$\pm$0.08 &  0.92 & -295.13 \\
1707 &H &  1243.583 &  1935.957 &   4.5390000 &   1.0104659 &    18.120 $\pm$   0.012 &    18.728 $\pm$   0.016 &    19.465 $\pm$   0.016 &    20.668 $\pm$   0.069  & -1.92$\pm$0.13 &  0.97 &         \\
1738 &N &   755.207 &  1964.317 &   4.5389328 &   1.0112524 &    18.922 $\pm$   0.029 &    19.490 $\pm$   0.043 &    20.148 $\pm$   0.019 &    21.201 $\pm$   0.259  & -1.99$\pm$0.39 &  0.97 &         \\
1744 &99 &  1238.353 &  1978.347 &   4.5388718 &   1.0104752 &    19.703 $\pm$   0.053 &    20.245 $\pm$   0.030 &    20.849 $\pm$   0.089 &    21.904 $\pm$   0.299  & -2.08$\pm$0.46 &  0.97 &         \\
1746 &131 &  1118.448 &  1977.566 &   4.5388784 &   1.0106683 &    19.244 $\pm$   0.024 &    19.737 $\pm$   0.033 &    20.324 $\pm$   0.038 &    21.283 $\pm$   0.078  & -1.60$\pm$0.37 &  0.97 &         \\
1766 &340 &  1496.295 &  1993.884 &   4.5388141 &   1.0100602 &    19.384 $\pm$   0.028 &    19.904 $\pm$   0.032 &    20.597 $\pm$   0.018 &    21.629 $\pm$   0.251  & -1.17$\pm$0.35 &  0.97 &         \\
1770 &97 &  1184.827 &  2003.372 &   4.5387983 &   1.0105619 &    19.366 $\pm$   0.074 &    19.896 $\pm$   0.040 &    20.566 $\pm$   0.024 &    21.609 $\pm$   0.112  & -1.48$\pm$0.58 &  0.95 &         \\
1772 &343 &  1557.547 &  1984.207 &   4.5388412 &   1.0099612 &    17.550 $\pm$   0.003 &    18.291 $\pm$   0.029 &    19.291 $\pm$   0.010 &    20.812 $\pm$   0.049  & -1.71$\pm$0.17 &  0.91 & -293.96 \\
1787 &116 &  1371.305 &  2021.150 &   4.5387368 &   1.0102620 &    19.128 $\pm$   0.037 &    19.715 $\pm$   0.031 &    20.466 $\pm$   0.024 &    21.538 $\pm$   0.202  & -1.61$\pm$0.30 &  0.97 &         \\
1788 &115 &  1348.013 &  2024.958 &   4.5387263 &   1.0102997 &    19.183 $\pm$   0.018 &    19.699 $\pm$   0.031 &    20.334 $\pm$   0.022 &    21.410 $\pm$   0.072  & -1.54$\pm$0.33 &  0.93 &         \\
1790 & &  1987.967 &  2023.971 &   4.5387034 &   1.0092688 &    18.045 $\pm$   0.007 &    18.710 $\pm$   0.015 &    19.573 $\pm$   0.013 &    20.840 $\pm$   0.079  & -1.75$\pm$0.11 &       &         \\
1791 &Q &   745.022 &  2031.946 &   4.5387282 &   1.0112703 &    19.553 $\pm$   0.051 &    20.078 $\pm$   0.040 &    20.650 $\pm$   0.027 &    21.647 $\pm$   0.206  & -2.11$\pm$0.45 &  0.89 &         \\
1807 &O &   695.720 &  2055.535 &   4.5386586 &   1.0113502 &    19.059 $\pm$   0.038 &    19.662 $\pm$   0.044 &    20.359 $\pm$   0.014 &    21.486 $\pm$   0.174  & -2.10$\pm$0.36 &  0.94 &         \\
\hline																 
\end{tabular}															 
\vfill																 
\end{minipage}															 
\end{sidewaystable*}

\addtocounter{table}{-1}

\begin{sidewaystable*}
\begin{minipage}[t][180mm]{\textwidth}
\caption{Continued.}
\label{FINALTAB}
\centering
\begin{tabular}{llrrlllllllllll}
\hline\hline             

INT ID & Other ID &  x & y & R.A. (2000) & Dec (2000) & $y\pm \epsilon_y$  &  $b\pm \epsilon_{b}$  & $v\pm \epsilon_v$ &  $u\pm \epsilon_u$ & [Fe/H]$_{\rm m_{\rm 1}}$ & P & $V_{\rm r}$\\

\hline
1813 &114 &  1338.130 &  2062.865 &   4.5386119 &   1.0103164 &    18.714 $\pm$   0.019 &    19.305 $\pm$   0.031 &    20.055 $\pm$   0.019 &    21.215 $\pm$   0.064  & -1.66$\pm$0.25 &  0.94 &         \\
1843 &437 &   282.485 &  1916.420 &   4.5390954 &   1.0120106 &    17.640 $\pm$   0.010 &    18.392 $\pm$   0.016 &    19.395 $\pm$   0.023 &    20.801 $\pm$   0.122  & -1.79$\pm$0.11 &  0.97 & -307.83 \\
1849 & &   370.640 &  2096.281 &   4.5385470 &   1.0118735 &    19.613 $\pm$   0.089 &    20.149 $\pm$   0.041 &    20.795 $\pm$   0.066 &    21.835 $\pm$   0.314  & -1.72$\pm$0.67 &  0.97 &         \\
1893 &335 &  1445.807 &  2144.639 &   4.5383596 &   1.0101447 &    18.062 $\pm$   0.009 &    18.726 $\pm$   0.024 &    19.639 $\pm$   0.014 &    21.010 $\pm$   0.039  & -1.45$\pm$0.17 &  0.96 & -288.59 \\
1900 & &  2027.494 &  2154.508 &   4.5383072 &   1.0092077 &    19.128 $\pm$   0.018 &    19.642 $\pm$   0.015 &    20.246 $\pm$   0.039 &    21.218 $\pm$   0.039  & -1.76$\pm$0.22 &       &         \\
1911 &55 &   858.762 &  2156.976 &   4.5383449 &   1.0110902 &    19.320 $\pm$   0.019 &    19.848 $\pm$   0.019 &    20.391 $\pm$   0.029 &    21.335 $\pm$   0.158  & -2.36$\pm$0.20 &  0.96 &         \\
1925 &M &  1097.671 &  2136.035 &   4.5383992 &   1.0107052 &    18.688 $\pm$   0.028 &    19.287 $\pm$   0.027 &    20.050 $\pm$   0.027 &    21.194 $\pm$   0.076  & -1.66$\pm$0.25 &  0.95 &         \\
1929 &108 &  1366.025 &  2175.235 &   4.5382705 &   1.0102739 &    19.241 $\pm$   0.020 &    19.791 $\pm$   0.036 &    20.506 $\pm$   0.048 &    21.584 $\pm$   0.110  & -1.42$\pm$0.34 &  0.90 &         \\
1933 & &  1864.355 &  2176.953 &   4.5382457 &   1.0094711 &    18.083 $\pm$   0.009 &    18.732 $\pm$   0.018 &    19.546 $\pm$   0.013 &    20.832 $\pm$   0.063  & -1.87$\pm$0.13 &       &         \\
1940 &11 &   495.329 &  2186.303 &   4.5382695 &   1.0116751 &    17.532 $\pm$   0.030 &    18.295 $\pm$   0.020 &    19.331 $\pm$   0.018 &    20.883 $\pm$   0.052  & -1.72$\pm$0.16 &  0.97 & -283.92 \\
1954 &49 &   764.593 &  2180.260 &   4.5382776 &   1.0112422 &    16.957 $\pm$   0.019 &    17.674 $\pm$   0.026 &    18.588 $\pm$   0.012 &    20.052 $\pm$   0.083  & -1.93$\pm$0.17 &  0.97 & -289.39 \\
1988 &249 &   267.862 &  2229.794 &   4.5381455 &   1.0120414 &    17.216 $\pm$   0.005 &    18.101 $\pm$   0.026 &    19.410 $\pm$   0.023 &    21.175 $\pm$   0.138  & -1.48$\pm$0.13 &  0.76 & -293.11 \\
2014 &22 &   641.982 &  2282.425 &   4.5379725 &   1.0114416 &    18.086 $\pm$   0.021 &    18.778 $\pm$   0.012 &    19.674 $\pm$   0.020 &    21.059 $\pm$   0.067  & -1.81$\pm$0.12 &  0.92 & -298.52 \\
2026 &322 &  1388.866 &  2298.269 &   4.5378971 &   1.0102396 &    18.813 $\pm$   0.007 &    19.407 $\pm$   0.031 &    20.157 $\pm$   0.027 &    21.355 $\pm$   0.086  & -1.68$\pm$0.24 &  0.92 &         \\
2027 &33 &   519.499 &  2298.652 &   4.5379281 &   1.0116388 &    18.681 $\pm$   0.024 &    19.206 $\pm$   0.021 &    19.771 $\pm$   0.024 &    20.775 $\pm$   0.194  & -2.18$\pm$0.23 &  0.95 &         \\
2058 &59 &   999.462 &  2323.009 &   4.5378361 &   1.0108672 &    19.213 $\pm$   0.046 &    19.781 $\pm$   0.034 &    20.528 $\pm$   0.024 &    21.708 $\pm$   0.265  & -1.41$\pm$0.39 &  0.96 &         \\
2082 &24 &   593.615 &  2347.761 &   4.5377760 &   1.0115207 &    17.073 $\pm$   0.023 &    17.929 $\pm$   0.022 &    19.029 $\pm$   0.017 &    20.713 $\pm$   0.098  & -2.12$\pm$0.12 &  0.92 & -273.35 \\
2086 &32 &   476.518 &  2285.816 &   4.5379682 &   1.0117077 &    18.933 $\pm$   0.071 &    19.550 $\pm$   0.029 &    20.276 $\pm$   0.031 &    21.417 $\pm$   0.199  & -2.07$\pm$0.38 &       &         \\
2090 &511 &  1636.014 &  2362.456 &   4.5376930 &   1.0098428 &    19.021 $\pm$   0.023 &    19.586 $\pm$   0.037 &    20.218 $\pm$   0.027 &    21.247 $\pm$   0.133  & -2.14$\pm$0.31 &  0.97 &         \\
2097 &45 &   661.980 &  2370.373 &   4.5377049 &   1.0114113 &    17.621 $\pm$   0.026 &    18.353 $\pm$   0.024 &    19.287 $\pm$   0.003 &    20.756 $\pm$   0.177  & -1.96$\pm$0.17 &  0.96 & -289.90 \\
2104 &70 &  1066.587 &  2354.007 &   4.5377398 &   1.0107599 &    15.899 $\pm$   0.007 &    16.452 $\pm$   0.015 &    17.408 $\pm$   0.004 &    18.688 $\pm$   0.019  &  0.19$\pm$0.16 &       &         \\
2106 &72 &  1093.260 &  2371.013 &   4.5376873 &   1.0107173 &    18.294 $\pm$   0.019 &    18.947 $\pm$   0.015 &    19.797 $\pm$   0.014 &    21.157 $\pm$   0.222  & -1.71$\pm$0.13 &  0.94 &         \\
2113 &324 &  1386.265 &  2382.759 &   4.5376410 &   1.0102456 &    18.560 $\pm$   0.006 &    19.120 $\pm$   0.024 &    19.795 $\pm$   0.021 &    20.903 $\pm$   0.069  & -1.80$\pm$0.20 &  0.97 &         \\
2119 &328 &  1491.485 &  2384.280 &   4.5376325 &   1.0100760 &    18.385 $\pm$   0.008 &    18.971 $\pm$   0.027 &    19.683 $\pm$   0.025 &    20.828 $\pm$   0.074  & -1.85$\pm$0.22 &  0.97 &         \\
2133 &317 &  1291.282 &  2403.164 &   4.5375829 &   1.0103990 &    19.525 $\pm$   0.038 &    20.031 $\pm$   0.024 &    20.629 $\pm$   0.026 &    21.607 $\pm$   0.229  & -1.69$\pm$0.34 &       &         \\
2143 & &   247.010 &  2411.917 &   4.5375934 &   1.0120790 &    18.424 $\pm$   0.041 &    19.044 $\pm$   0.050 &    19.833 $\pm$   0.004 &    21.056 $\pm$   0.107  & -1.73$\pm$0.40 &  0.97 &         \\
2149 &K &  1233.863 &  2414.468 &   4.5375504 &   1.0104917 &    18.125 $\pm$   0.011 &    18.783 $\pm$   0.020 &    19.661 $\pm$   0.011 &    21.013 $\pm$   0.143  & -1.60$\pm$0.14 &  0.95 &         \\
2192 &41 &   789.659 &  2475.832 &   4.5373802 &   1.0112081 &    18.271 $\pm$   0.015 &    18.915 $\pm$   0.032 &    19.727 $\pm$   0.022 &    21.047 $\pm$   0.056  & -1.83$\pm$0.22 &  0.97 &         \\
2194 &506 &  1628.285 &  2475.235 &   4.5373516 &   1.0098574 &    17.914 $\pm$   0.007 &    18.571 $\pm$   0.025 &    19.386 $\pm$   0.027 &    20.697 $\pm$   0.073  & -1.94$\pm$0.18 &  0.95 & -295.22 \\
2206 &505 &  1589.615 &  2495.234 &   4.5372925 &   1.0099201 &    19.696 $\pm$   0.034 &    20.212 $\pm$   0.049 &    20.803 $\pm$   0.055 &    21.797 $\pm$   0.140  & -1.88$\pm$0.50 &  0.82 &         \\
2226 &449 &   243.181 &  2521.552 &   4.5372605 &   1.0120873 &    17.487 $\pm$   0.012 &    18.255 $\pm$   0.039 &    19.301 $\pm$   0.031 &    20.886 $\pm$   0.130  & -1.71$\pm$0.22 &  0.97 & -305.70 \\
2243 & &   329.176 &  2548.725 &   4.5371752 &   1.0119499 &    19.868 $\pm$   0.024 &    20.380 $\pm$   0.048 &    20.939 $\pm$   0.018 &    21.870 $\pm$   0.356  & -2.06$\pm$0.45 &       &         \\
2244 &305 &  1080.267 &  2549.078 &   4.5371475 &   1.0107418 &    19.250 $\pm$   0.028 &    19.794 $\pm$   0.022 &    20.532 $\pm$   0.030 &    21.597 $\pm$   0.156  & -1.18$\pm$0.28 &  0.94 &         \\
2245 &290 &   960.225 &  2549.427 &   4.5371509 &   1.0109351 &    19.272 $\pm$   0.016 &    19.836 $\pm$   0.034 &    20.549 $\pm$   0.026 &    21.582 $\pm$   0.106  & -1.59$\pm$0.30 &  0.87 &         \\
2247 &312 &  1336.335 &  2546.478 &   4.5371466 &   1.0103292 &    18.823 $\pm$   0.034 &    19.355 $\pm$   0.020 &    19.996 $\pm$   0.011 &    21.089 $\pm$   0.174  & -1.70$\pm$0.26 &  0.94 &         \\
2252 &273 &   717.847 &  2558.424 &   4.5371323 &   1.0113252 &    18.241 $\pm$   0.025 &    18.888 $\pm$   0.051 &    19.744 $\pm$   0.020 &    21.049 $\pm$   0.096  & -1.60$\pm$0.38 &  0.95 & -276.62 \\
2256 &S-430 &  1741.733 &  2562.882 &   4.5370822 &   1.0096762 &    18.206 $\pm$   0.022 &    18.805 $\pm$   0.034 &    19.572 $\pm$   0.045 &    20.716 $\pm$   0.098  & -1.65$\pm$0.29 &       &         \\
2261 & &    35.577 &  2571.963 &   4.5371146 &   1.0124215 &    19.161 $\pm$   0.027 &    19.696 $\pm$   0.062 &    20.314 $\pm$   0.013 &    21.318 $\pm$   0.194  & -1.90$\pm$0.56 &       &         \\
2309 & &   129.057 &  2640.499 &   4.5369039 &   1.0122730 &    19.472 $\pm$   0.027 &    20.016 $\pm$   0.033 &    20.632 $\pm$   0.014 &    21.690 $\pm$   0.126  & -2.02$\pm$0.30 &       &         \\
2321 & &  1119.116 &  2656.535 &   4.5368204 &   1.0106813 &    18.849 $\pm$   0.006 &    19.420 $\pm$   0.026 &    20.094 $\pm$   0.012 &    21.249 $\pm$   0.159  & -1.93$\pm$0.20 &  0.97 &         \\
2334 &286 &   979.283 &  2672.938 &   4.5367751 &   1.0109068 &    17.693 $\pm$   0.005 &    18.433 $\pm$   0.021 &    19.486 $\pm$   0.012 &    21.019 $\pm$   0.143  & -1.45$\pm$0.13 &  0.97 & -301.91 \\
2340 & &   595.716 &  2689.268 &   4.5367389 &   1.0115243 &    19.473 $\pm$   0.047 &    19.987 $\pm$   0.055 &    20.639 $\pm$   0.032 &    21.655 $\pm$   0.109  & -1.39$\pm$0.62 &  0.96 &         \\
2346 &302 &  1207.025 &  2692.480 &   4.5367084 &   1.0105404 &    19.680 $\pm$   0.042 &    20.174 $\pm$   0.011 &    20.776 $\pm$   0.033 &    21.742 $\pm$   0.141  & -1.50$\pm$0.31 &  0.94 &         \\
\hline																 
\end{tabular}															 
\vfill																 
\end{minipage}															 
\end{sidewaystable*}

\addtocounter{table}{-1}

\begin{sidewaystable*}
\begin{minipage}[t][180mm]{\textwidth}
\caption{Continued.}
\label{FINALTAB}
\centering
\begin{tabular}{llrrlllllllllll}
\hline\hline             

INT ID & Other ID &  x & y & R.A. (2000) & Dec (2000) & $y\pm \epsilon_y$  &  $b\pm \epsilon_{b}$  & $v\pm \epsilon_v$ &  $u\pm \epsilon_u$ & [Fe/H]$_{\rm m_{\rm 1}}$ & P & $V_{\rm r}$\\

\hline
2366 &267 &   549.346 &  2687.772 &   4.5367455 &   1.0115988 &    17.073 $\pm$   0.014 &    17.977 $\pm$   0.030 &    19.298 $\pm$   0.043 &    21.196 $\pm$   0.244  & -1.56$\pm$0.16 &  0.95 & -291.89 \\
2375 &500 &  1459.778 &  2714.014 &   4.5366340 &   1.0101335 &    18.747 $\pm$   0.019 &    19.277 $\pm$   0.041 &    19.913 $\pm$   0.020 &    20.914 $\pm$   0.062  & -1.71$\pm$0.38 &  0.96 &         \\
2381 &281 &   741.638 &  2715.270 &   4.5366554 &   1.0112901 &    17.970 $\pm$   0.011 &    18.641 $\pm$   0.029 &    19.579 $\pm$   0.022 &    21.000 $\pm$   0.058  & -1.40$\pm$0.22 &  0.97 & -274.95 \\
2409 &S-172 &  1506.801 &  2745.654 &   4.5365367 &   1.0100583 &    16.813 $\pm$   0.014 &    17.582 $\pm$   0.015 &    18.577 $\pm$   0.014 &    20.144 $\pm$   0.038  & -1.96$\pm$0.09 &       &         \\
2421 &297 &  1033.593 &  2768.154 &   4.5364842 &   1.0108211 &    17.811 $\pm$   0.012 &    18.517 $\pm$   0.022 &    19.490 $\pm$   0.008 &    21.025 $\pm$   0.283  & -1.54$\pm$0.14 &  0.97 & -287.74 \\
2428 & &   533.354 &  2772.753 &   4.5364876 &   1.0116261 &    18.549 $\pm$   0.024 &    19.147 $\pm$   0.044 &    19.921 $\pm$   0.023 &    21.082 $\pm$   0.081  & -1.58$\pm$0.36 &  0.97 &         \\
2441 &490 &  1129.449 &  2795.155 &   4.5363994 &   1.0106672 &    17.534 $\pm$   0.009 &    18.280 $\pm$   0.029 &    19.341 $\pm$   0.008 &    20.878 $\pm$   0.120  & -1.46$\pm$0.18 &  0.97 & -303.89 \\
2443 &488 &  1100.876 &  2813.530 &   4.5363445 &   1.0107136 &    19.901 $\pm$   0.042 &    20.421 $\pm$   0.049 &    21.039 $\pm$   0.049 &    22.047 $\pm$   0.168  & -1.73$\pm$0.54 &  0.90 &         \\
2457 &S-158 &  1903.162 &  2830.568 &   4.5362649 &   1.0094210 &    18.675 $\pm$   0.021 &    19.200 $\pm$   0.015 &    19.836 $\pm$   0.009 &    20.892 $\pm$   0.054  & -1.65$\pm$0.18 &       &         \\
2462 & &   478.088 &  2835.226 &   4.5363002 &   1.0117161 &    18.756 $\pm$   0.019 &    19.272 $\pm$   0.024 &    19.890 $\pm$   0.023 &    20.852 $\pm$   0.101  & -1.68$\pm$0.26 &  0.95 &         \\
2480 & &  1232.126 &  2836.178 &   4.5362716 &   1.0105026 &    19.307 $\pm$   0.007 &    19.908 $\pm$   0.038 &    20.632 $\pm$   0.021 &    21.712 $\pm$   0.295  & -1.92$\pm$0.28 &  0.82 &         \\
2493 &S-151 &  1672.042 &  2876.774 &   4.5361333 &   1.0097944 &    17.286 $\pm$   0.012 &    18.089 $\pm$   0.018 &    19.259 $\pm$   0.010 &    20.896 $\pm$   0.073  & -1.43$\pm$0.11 &       &         \\
2501 &473 &   756.903 &  2894.155 &   4.5361118 &   1.0112689 &    17.465 $\pm$   0.018 &    18.240 $\pm$   0.039 &    19.387 $\pm$   0.010 &    21.022 $\pm$   0.090  & -1.32$\pm$0.23 &  0.93 & -289.79 \\
2506 & &  1230.702 &  2895.587 &   4.5360913 &   1.0105059 &    18.559 $\pm$   0.009 &    19.175 $\pm$   0.036 &    19.957 $\pm$   0.029 &    21.241 $\pm$   0.127  & -1.73$\pm$0.27 &       &         \\
2546 & &    26.218 &  2946.628 &   4.5359769 &   1.0124437 &    19.059 $\pm$   0.021 &    19.608 $\pm$   0.039 &    20.233 $\pm$   0.031 &    21.284 $\pm$   0.107  & -2.02$\pm$0.35 &       &         \\
2552 & &  1961.773 &  2957.478 &   4.5358787 &   1.0093287 &    18.743 $\pm$   0.024 &    19.345 $\pm$   0.021 &    20.089 $\pm$   0.047 &    21.269 $\pm$   0.197  & -1.81$\pm$0.24 &       &         \\
2627 & &  1100.652 &  3042.528 &   4.5356498 &   1.0107180 &    19.379 $\pm$   0.018 &    19.913 $\pm$   0.030 &    20.545 $\pm$   0.030 &    21.527 $\pm$   0.084  & -1.79$\pm$0.28 &       &         \\
2694 & &   871.089 &  3114.065 &   4.5354400 &   1.0110888 &    18.875 $\pm$   0.016 &    19.457 $\pm$   0.023 &    20.180 $\pm$   0.026 &    21.369 $\pm$   0.292  & -1.73$\pm$0.21 &       &         \\
2735 &S-98 &   974.017 &  3171.142 &   4.5352635 &   1.0109241 &    19.483 $\pm$   0.049 &    19.992 $\pm$   0.033 &    20.589 $\pm$   0.043 &    21.582 $\pm$   0.194  & -1.74$\pm$0.48 &       &         \\
2739 &S-419 &  1508.020 &  3165.365 &   4.5352635 &   1.0100636 &    18.068 $\pm$   0.021 &    18.709 $\pm$   0.030 &    19.507 $\pm$   0.025 &    20.735 $\pm$   0.075  & -1.89$\pm$0.22 &       &         \\
2772 &S-91 &  1891.718 &  3205.242 &   4.5351305 &   1.0094458 &    18.828 $\pm$   0.010 &    19.421 $\pm$   0.030 &    20.164 $\pm$   0.049 &    21.270 $\pm$   0.143  & -1.73$\pm$0.27 &       &         \\
2775 & &   935.719 &  3205.964 &   4.5351591 &   1.0109863 &    18.408 $\pm$   0.011 &    18.972 $\pm$   0.037 &    19.694 $\pm$   0.027 &    20.871 $\pm$   0.039  & -1.53$\pm$0.31 &       &         \\
2845 & &   175.214 &  3283.989 &   4.5349469 &   1.0122101 &    18.299 $\pm$   0.018 &    18.922 $\pm$   0.028 &    19.699 $\pm$   0.015 &    20.988 $\pm$   0.115  & -1.83$\pm$0.22 &       &         \\
2867 &S-74  &  1741.196 &  3300.511 &   4.5348468 &   1.0096899 &    18.809 $\pm$   0.027 &    19.379 $\pm$   0.032 &    20.050 $\pm$   0.047 &    21.154 $\pm$   0.103  & -1.94$\pm$0.31 &       &         \\
2877 & &   231.718 &  3319.647 &   4.5348363 &   1.0121200 &    18.428 $\pm$   0.026 &    19.026 $\pm$   0.021 &    19.820 $\pm$   0.011 &    21.057 $\pm$   0.083  & -1.47$\pm$0.21 &       &         \\
2891 &S-413 &  1113.035 &  3331.541 &   4.5347724 &   1.0107027 &    17.985 $\pm$   0.011 &    18.653 $\pm$   0.029 &    19.574 $\pm$   0.021 &    20.948 $\pm$   0.118  & -1.46$\pm$0.20 &       &         \\
2899 & &   460.810 &  3362.415 &   4.5346994 &   1.0117526 &    19.109 $\pm$   0.024 &    19.673 $\pm$   0.045 &    20.332 $\pm$   0.037 &    21.412 $\pm$   0.168  & -1.95$\pm$0.39 &       &         \\
2921 & &   693.073 &  3400.754 &   4.5345755 &   1.0113797 &    18.765 $\pm$   0.012 &    19.340 $\pm$   0.023 &    20.015 $\pm$   0.037 &    21.095 $\pm$   0.128  & -1.97$\pm$0.21 &       &         \\
2932 &S-60 &  1586.930 &  3416.387 &   4.5345006 &   1.0099405 &    18.994 $\pm$   0.017 &    19.557 $\pm$   0.022 &    20.210 $\pm$   0.018 &    21.278 $\pm$   0.081  & -1.98$\pm$0.19 &       &         \\
2939 &S-58 &  1471.657 &  3427.579 &   4.5344701 &   1.0101264 &    17.371 $\pm$   0.014 &    18.165 $\pm$   0.027 &    19.324 $\pm$   0.007 &    20.997 $\pm$   0.036  & -1.41$\pm$0.16 &       &         \\
2964 &S-53 &  1126.982 &  3469.216 &   4.5343547 &   1.0106823 &    18.682 $\pm$   0.013 &    19.204 $\pm$   0.026 &    19.819 $\pm$   0.012 &    20.775 $\pm$   0.110  & -1.77$\pm$0.24 &       &         \\
2996 &S-47 &  1098.640 &  3507.532 &   4.5342388 &   1.0107286 &    18.871 $\pm$   0.007 &    19.445 $\pm$   0.021 &    20.144 $\pm$   0.039 &    21.193 $\pm$   0.116  & -1.80$\pm$0.20 &       &         \\
2999 & &   255.217 &  3503.146 &   4.5342789 &   1.0120848 &    17.106 $\pm$   0.032 &    17.933 $\pm$   0.033 &    19.104 $\pm$   0.012 &    20.760 $\pm$   0.086  & -1.62$\pm$0.20 &       &         \\
3003 & &   637.980 &  3522.039 &   4.5342093 &   1.0114701 &    19.434 $\pm$   0.025 &    19.949 $\pm$   0.054 &    20.571 $\pm$   0.064 &    21.540 $\pm$   0.192  & -1.63$\pm$0.59 &       &         \\
3029 & &   517.222 &  3560.259 &   4.5340967 &   1.0116647 &    18.411 $\pm$   0.011 &    18.939 $\pm$   0.027 &    19.564 $\pm$   0.008 &    20.565 $\pm$   0.067  & -1.76$\pm$0.25 &       &         \\
3034 & &    67.449 &  3579.026 &   4.5340543 &   1.0123872 &    18.928 $\pm$   0.034 &    19.458 $\pm$   0.023 &    20.096 $\pm$   0.011 &    21.199 $\pm$   0.145  & -1.70$\pm$0.29 &       &         \\
3068 &S-37 &  1218.038 &  3619.460 &   4.5338960 &   1.0105377 &    18.818 $\pm$   0.021 &    19.369 $\pm$   0.048 &    20.002 $\pm$   0.012 &    21.069 $\pm$   0.101  & -1.98$\pm$0.40 &       &         \\
3075 &S-34 &  1247.637 &  3637.357 &   4.5338407 &   1.0104904 &    19.291 $\pm$   0.015 &    19.851 $\pm$   0.048 &    20.525 $\pm$   0.037 &    21.505 $\pm$   0.155  & -1.81$\pm$0.42 &       &         \\
3139 &S-19 &  1150.251 &  3780.995 &   4.5334082 &   1.0106492 &    18.715 $\pm$   0.021 &    19.290 $\pm$   0.013 &    19.971 $\pm$   0.015 &    21.064 $\pm$   0.030  & -1.94$\pm$0.15 &       &         \\
3163 &S-14 &   894.363 &  3810.251 &   4.5333271 &   1.0110615 &    17.704 $\pm$   0.011 &    18.395 $\pm$   0.023 &    19.308 $\pm$   0.013 &    20.706 $\pm$   0.026  & -1.72$\pm$0.15 &       &         \\
3165 &S-11 &  1171.659 &  3811.070 &   4.5333166 &   1.0106151 &    17.644 $\pm$   0.004 &    18.359 $\pm$   0.021 &    19.287 $\pm$   0.012 &    20.727 $\pm$   0.046  & -1.85$\pm$0.13 &       &         \\
3221 & &   863.235 &  3898.302 &   4.5330606 &   1.0111127 &    19.492 $\pm$   0.051 &    19.990 $\pm$   0.022 &    20.675 $\pm$   0.065 &    21.728 $\pm$   0.181  & -0.91$\pm$0.51 &       &         \\
3228 & &   413.595 &  3917.286 &   4.5330167 &   1.0118357 &    18.569 $\pm$   0.017 &    19.132 $\pm$   0.015 &    19.808 $\pm$   0.014 &    20.956 $\pm$   0.074  & -1.84$\pm$0.14 &       &         \\
3233 & &   975.794 &  3920.596 &   4.5329900 &   1.0109318 &    18.723 $\pm$   0.018 &    19.289 $\pm$   0.021 &    19.964 $\pm$   0.034 &    21.031 $\pm$   0.070  & -1.87$\pm$0.21 &       &         \\
													 
\hline																 
\end{tabular}															 
\vfill																 
\end{minipage}															 
\end{sidewaystable*}

\end{document}